\documentclass[12pt]{article}

\usepackage[T1]{fontenc} 				

\usepackage{amssymb,amsmath,amscd}
\usepackage[english]{babel}
\usepackage{titlesec}
\usepackage{slashed}
\usepackage{hyperref}
\usepackage{graphicx,xcolor}
\usepackage{enumerate}
\usepackage{bbm} 						
\usepackage{float}
\usepackage{mathtools}
\usepackage{subfig}
\hypersetup{
	colorlinks=true,
	linkcolor=black,
	citecolor=dark-red,
	urlcolor=dark-green,
	linktoc=all
}

\usepackage{geometry} 					
\numberwithin{equation}{section} 

\titleformat{\section}[block]{\Large\bfseries\centering}{\thesection}{1em}{} 
\titleformat{\subsection}[block]{\bfseries}{\thesubsection}{1em}{} 

\definecolor{dark-gray}{gray}{0.20}
\definecolor{gray}{gray}{0.30}
\definecolor{light-gray}{gray}{0.80}
\definecolor{dark-red}{rgb}{0.7,0,0}
\definecolor{dark-green}{rgb}{0.1,0.4,0}
\definecolor{dark-blue}{rgb}{0.3,0.3,0.7}
\definecolor{light-blue}{rgb}{0.8,0.8,1}
\definecolor{cardinal}{rgb}{0.6,0,0}
\definecolor{darkgreen}{rgb}{0,0.5,0}
\definecolor{golden}{rgb}{0.92, 0.7, 0}
\definecolor{midnight}{rgb}{0, 0, 0.5}
\definecolor{darkblue}{rgb}{0.2, 0, 0.8}
\definecolor{forestgreen}{rgb}{0.13, 0.55, 0.13}

\def\cD{{\cal D}}
\def\cF{{\cal F}}

\def\cN{{\cal N}}

\def\cZ{{\cal Z}}

\newcommand{\Dx}{\mathsf{D}}

\def\SU{{\rm SU}}

\title{\fontsize{20pt}{24pt}{The planar limit of the $\mathcal{N} = 2$ $\mathbf{E}$-theory: numerical calculations and the large $\lambda$ expansion\\
	}\vspace{3mm}}

\bigskip
\bigskip

\author{Nikolay Bobev, Pieter-Jan De Smet, and Xuao Zhang\\[5mm]
	\normalsize Instituut voor Theoretische Fysica, K.U. Leuven\\
	\normalsize Celestijnenlaan 200D, BE-3001 Leuven, Belgium\\[2mm]
	\texttt{\small\href{mailto:nikolay.bobev@kuleuven.be}{nikolay.bobev@kuleuven.be}, \href{mailto::pieterjan.desmet@kuleuven.be}{pieterjan.desmet@kuleuven.be}, \href{mailto::xuao.zhang@kuleuven.be}{xuao.zhang@kuleuven.be}}\\
}

\date{}

\begin{document}  
	
\maketitle
\begin{abstract}
	\noindent We study correlation functions of local operators and Wilson loop expectation values in the planar limit of a 4d $\mathcal{N}=2$ superconformal $\SU(N)$ YM theory with hypermultiplets in the symmetric and antisymmetric representations of the gauge group. This so called $\mathbf{E}$ theory is closely related to $\mathcal{N}=4$ SYM and has a holographic description in terms of a $\mathbb{Z}_2$ orientifold of AdS$_5\times S^5$. Using recent matrix model results based on supersymmetric localization we develop efficient numerical methods to calculate two- and three-point functions of certain single trace operators as well as 1/2-BPS Wilson loop expectation values as a function of the 't Hooft coupling $\lambda$. We use our numerical results to arrive at simple analytic expressions for these correlators valid up to sixth order in the $\lambda^{-1/2}$ strong coupling expansion. These results provide explicit field theory predictions for the $\alpha'$ corrections to the supergravity approximation of type IIB string theory on the AdS$_5\times S^5/\mathbb{Z}_2$ orientifold.
\end{abstract}

\clearpage

\tableofcontents


\section{Introduction and summary of results}

Since the seminal work of 't Hooft \cite{tHooft:1973alw} the planar limit of four-dimensional gauge theories has served as an important approximation that provides new insights into their dynamics. In subsequent developments a plethora of additional tools, amongst which integrability, supersymmetric localization, and AdS/CFT, were applied with great efficacy in conjunction to the planar limit to improve our understanding of gauge and string theory. A central example which often serves as a testbed for new calculational techniques and conceptual advances is the $\mathcal{N}=4$ SYM theory. Our modest goal in this work is to build on recent results from supersymmetric localization for four-dimensional $\mathcal{N}=2$ SCFTs, see \cite{Pestun:2016zxk} for a review, and derive some new results for physical observables in a particular planar gauge theory that shares many properties with the $\mathcal{N}=4$ SYM theory.

The theory of interest here is an $\mathcal{N}=2$ $\SU(N)$ gauge theory with one hypermultiplet in the symmetric representation of the gauge group and one hypermultiplet in the antisymmetric representation. This was referred to as the $\mathbf{E}$ theory in \cite{Billo1} and we will also adopt this monicker. As discussed in \cite{Billo1} the $\mathbf{E}$ theory is closely related to $\SU(N)$ $\mathcal{N}=4$ SYM theory. For instance, the conformal anomaly coefficients of the two theories are
\begin{equation}\label{eq:acEN4}
a^{\mathcal{N}=4}=c^{\mathcal{N}=4}=\frac{1}{4}(N^2-1)\,, \qquad a^{\mathbf{E}} = a^{\mathcal{N}=4} + \frac{1}{24}\,, \qquad c^{\mathbf{E}} = c^{\mathcal{N}=4} + \frac{1}{12}\,.
\end{equation}
Moreover, similarly to $\mathcal{N}=4$ SYM, the $\mathbf{E}$ theory theory has a holographic description in terms of type IIB string theory on a $\mathbb{Z}_2$ orientifold of AdS$_5\times S^5$, see \cite{Park:1999ep,Ennes:2000fu}. 

Certain physical observables in both the $\mathbf{E}$ theory and $\mathcal{N}=4$ SYM can be calculated by placing the theory on $S^4$ and employing supersymmetric localization as in \cite{Pestun:2007rz}. Importantly, in the planar limit the instanton contributions to the supersymmetric localization matrix model trivialize and the calculation of some physical observables as a function of the 't Hooft coupling $\lambda$ becomes feasible. This in turn provides the interesting possibility for explicit gauge theory calculations in the strong coupling limit, $\lambda \gg 1$, which should lead to insights into the $\alpha'$ corrections of type IIB string theory around the AdS$_5\times S^5/ \mathbb{Z}_2$ orientifold. Obtaining such strong coupling results in the planar limit of the $\mathbf{E}$ theory is the main objective of this work. 

In particular, we will be mainly interested in the two-point correlation functions of single trace operators in the $\mathbf{E}$ theory built out of the complex scalar field in the $\mathcal{N}=2$ vector multiplet. Conformal covariance and the R-symmetry of the theory dictate that these two-point functions take the following form
\begin{equation}
	\langle {\rm tr}\varphi^m(x) {\rm tr}\bar{\varphi}^n(0) \rangle = \frac{G_m(\lambda,N)\delta_{m,n}}{(4\pi^2 x^2)^{2m}}\,,
\end{equation}
where $G_m(\lambda,N)$ is a non-trivial function that can in principle be computed by supersymmetric localization, however explicit calculations for general values of $N$ are difficult. To organize the calculation it is useful to take the ratio of this function to the corresponding two-point function in $\mathcal{N}=4$ SYM, $G_m^{(0)}$. Moreover, it is important to distinguish the operators with $m$ even and odd in the planar limit. The operators with odd $m$ correspond to twisted sector modes in the AdS$_5\times S^5/ \mathbb{Z}_2$ orientifold and have distinct correlators from those of $\mathcal{N}=4$ SYM already at leading order in the planar limit. The untwisted sector modes correspond to operators with even $m$ and their two-point functions are identical with those in $\mathcal{N}=4$ SYM to leading order at large $N$ but differ at subleading orders. This information can be conveniently organized into the following large $N$ expansions 
\begin{equation}
	\frac{G_{2k+1}}{G_{2k+1}^{(0)}} = 1 + \Delta_k(\lambda) + O(N^{-2})\,,\qquad 	\frac{G_{2k}}{G_{2k}^{(0)}} = 1 + \frac{\delta_k(\lambda)}{N^2} + O(N^{-4})\,.
	\label{DeltaAnddelta}
\end{equation}
As explained in \cite{Billo1}, $\Delta_k(\lambda)$ and $\delta_k(\lambda)$ can be calculated by using supersymmetric localization and matrix model techniques. It was shown in \cite{Billo1} that a central role in this calculation is played by an infinite matrix $\mathsf{X}$ with matrix elements:
\begin{align} \label{Xx}
	\mathsf{X}_{kl} = (-1)^{k+l+1}8 \sqrt{(2k+1)(2l+1)} \int_0^\infty \! dt\ W(t) 
	J_{2k+1}\Big(\frac{t\sqrt{\lambda}}{2\pi}\Big)\, 
	J_{2l+1}\Big(\frac{t\sqrt{\lambda}}{2\pi}\Big)\,,	
\end{align}
with $k,l = 1 , 2 , \ldots$, where $J_n$ are Bessel functions, and 
\begin{equation} \label{defW}
	W(t) =\frac{e^t}{t (e^t-1)^2} \,.
\end{equation}
The calculation of $\Delta_k(\lambda)$ then amounts to inverting $1-\mathsf{X}$ and calculating the following ratio of determinants\footnote{Here the symbol $\Dx_{(k)}$ is used to denote the upper left $k \times k$ block of the matrix $\Dx$. }
\begin{align}
	\label{defDeltaINTRO}
	1 + \Delta_k(\lambda) = 
	\frac{\det \Dx_{(k)}}{\det \Dx_{(k-1)}},\qquad \Dx_{k,l}\equiv \Big(\frac{1}{1 - \mathsf{X}}\Big)_{k,l}\,.
\end{align} 		
It is relatively easy to calculate the small $\lambda$ expansion of $\Delta_k(\lambda)$ using~\eqref{defDeltaINTRO}. Indeed, for each order in the small $\lambda$ expansion, the matrix $\mathsf{X}$ is essentially a finite matrix, and the calculation of $\Delta_k(\lambda)$ order by order in the small $\lambda$ limit amounts to straightforward calculations with finite matrices. However, it is more challenging to calculate the large $\lambda$ expansion of $\Delta_k(\lambda)$. To tackle this problem we develop numerical techniques based on solving an integral equation that allow for the efficient calculation of $\Delta_k(\lambda)$ for general finite values of $\lambda$. Using this method for large values of $\lambda$ and performing a precise numerical fitting procedure allows us to propose a conjecture for the analytic form of the first six terms in the large $\lambda$ expansion of $\Delta_k(\lambda)$.  The first three terms in this expansion read
\begin{equation}\label{eq:NikolayINTRO}
1+\Delta_k(\lambda) = 8 \pi^2 k (2 k+1) \left[ \frac{1}{\lambda}-\frac{16 k \log 2}{\lambda^{3/2}}
+\frac{32 k (4 k -1)\log^2 2}{\lambda^{2}}\right]+O(\lambda^{-5/2})\,,
\end{equation}
while three more subleading terms are presented in Section~\ref{sec:Deltamoreterms}. The leading $1/\lambda$ term in \eqref{eq:NikolayINTRO} agrees with the analysis in \cite{Billo1} where it was calculated using analytic methods. We have also attempted to provide an analytic derivation of \eqref{eq:NikolayINTRO} but we were not successful due to certain subtleties we encountered when working with infinite matrices. This attempt is summarized in Appendix~\ref{sec:expansionDeltak}.

It was argued in \cite{Billo1} that $\delta_k(\lambda)$ in \eqref{DeltaAnddelta} can be calculated by taking derivatives with respect to $\lambda$ of the difference between the $S^4$ free energy of the $\mathbf{E}$ theory and $\mathcal{N}=4$ SYM. It was shown in \cite{Tseytlin} that this difference in free energies is related to the determinant of the matrix $\mathsf{X}$ as follows: 
\begin{equation}
	{\cal F} \equiv F^{\rm E-theory} - F^{\cN=4} = \frac{1}{2} \log \det ( 1 - \mathsf{X})\,.
\end{equation}
Again, it is relatively straightforward to calculate the small $\lambda$ expansion of ${\cal F}$, but it is much more difficult to calculate its large $\lambda$ expansion. To this end we develop a numerical method to compute the determinant of $1 - \mathsf{X}$ for general values of $\lambda$. Detailed numerical analysis in the large $\lambda$ regime leads us to conjecture the following analytic form of the first four terms in the large $\lambda$ expansion of $\mathcal{F}$
\begin{equation}\label{eq:Flargelambdaintro}
{\cal F} = \frac{1}{8} \sqrt{\lambda}- \frac{3}{8} \log \left(\frac{\lambda}{\lambda_0}\right) - \frac{3}{4} \frac{\log 2}{\sqrt{\lambda}} 
- \frac{3}{2} \frac{\log^2 2}{\lambda} + O( \lambda^{-3/2})\,,
\end{equation}
where $\lambda_0 \approx 7.72390117$ is a numerical constant that we were not able to identify in terms of familiar irrational or transcendental numbers.\footnote{See the discussion around \eqref{eq:lambda0v2} added in v2 of this work.} We note that the $1/8$ coefficient of the leading $\sqrt{\lambda}$ term in \eqref{eq:Flargelambdaintro} differs from the result in \cite{Tseytlin} where analytical methods were used to find the value $1/2\pi$ for this coefficient. We have no reason to doubt our very precise numerical analysis, which also allows us to find three more subleading terms in the large $\lambda$ expansion as compared to \cite{Tseytlin}, and believe that the discrepancy may be due to subtleties with the analytic method used in \cite{Tseytlin} when applied to matrices of infinite rank.
The result for $\mathcal{F}$ in \eqref{eq:Flargelambdaintro} can be combined with the arguments in \cite{Billo1} to find the large $\lambda$ expansion of the two-point function $\delta_{k}(\lambda)$. We find the following explicit result for the leading four terms in the large $\lambda$ expansion
\begin{multline}
\delta_k(\lambda) =  -\frac{k(4k^2-1) }{16}\sqrt{\lambda} + \frac{3}{8}k(4k^2-2)\\- \frac{3}{8}\frac{k(4k^2-3)\log 2}{\sqrt{\lambda}} -\frac{3}{2}  \frac{k(4k^2-4) \log^22}{\lambda}+O(\lambda^{-3/2})\,.
\end{multline}

Knowing the difference of free energies $\mathcal{F}$ allows us also to calculate the vacuum expectation value of a circular supersymmetric Wilson loop wrapping the equator of $S^4$. For $\mathcal{N}=4$ SYM in the planar limit this Wilson loop vev is given in terms of a Bessel function, see \cite{Drukker:2000rr}, 
\begin{equation}
\langle W^{\mathcal{N}=4}\rangle_0 = \frac{2N}{\sqrt{\lambda}} I_1(\sqrt{\lambda})\,.
\end{equation}
The deviations from this leading planar result are captured by a function $q(\lambda)$, see \cite{Tseytlin}
\begin{equation}\label{eq:Deltaqdef}
\frac{\langle W^{\mathcal{N}=4}\rangle}{\langle W^{\mathcal{N}=4}\rangle_0} = 1 + \frac{q^{\mathcal{N}=4}(\lambda)}{N^2} + O(N^{-4})\,, ~~ \text{where} ~~ q^{\mathcal{N}=4}(\lambda) = \frac{\lambda^{3/2}}{96} \frac{I_2(\sqrt{\lambda})}{I_1(\sqrt{\lambda})} - \frac{\lambda}{8}\,.
\end{equation}
As discussed in \cite{Tseytlin} for the $\mathbf{E}$ theory one finds that the leading order result for the Wilson loop vev in the planar limit is the same as that for $\mathcal{N}=4$ SYM, $\langle W^{\mathbf{E}}\rangle_0  = \langle W^{\mathcal{N}=4}\rangle_0 $. There are differences however at order $N^{-2}$ which can be captured by the function~$\Delta q(\lambda) = q^{\mathbf{E}}(\lambda)-q^{\mathcal{N}=4}(\lambda)$. It was furthermore shown in \cite{Tseytlin} that $\Delta q(\lambda) = -\frac{\lambda^2}{4}\partial_\lambda \mathcal{F}$. Using this relation and the result in \eqref{eq:Flargelambdaintro} we find the following large $\lambda$ behavior
\begin{equation}\label{eq:Deltaqres}
\Delta q (\lambda) = - \frac{1}{64} \lambda^{3/2}+\frac{3}{32} \lambda - \frac{3\log2}{32} \lambda^{1/2} - \frac{3\log^22}{8} +O(\lambda^{-1/2})\,.
\end{equation}

Our results for the large $\lambda$ behavior of $\mathcal{F}$ and $\Dx_{k,l}$ can be used in conjunction with the recent studies in \cite{Billo:2022xas} to derive the strong coupling behavior of the planar limit of some three-point extremal correlators of single trace operators in the $\mathbf{E}$ theory. We are able to compute these correlations functions up to order $\lambda^{-2}$ in the strong coupling expansion, which improves on the results of \cite{Billo:2022xas} by providing three additional terms in the large $\lambda$ expansion. These results are presented in Section~\ref{sec:Threeptfc}.

We stress that our numerical methods to calculate  $\Delta_k(\lambda)$ and  ${\cal F}$ are useful independently of the large $\lambda$ results presented above. The numerical method we propose is fast and efficient and provides accurate results for a wide range of values for the coupling  $\lambda$, $0 \le  \lambda \lesssim 10^7 $, for both $\Delta_k(\lambda)$ and  ${\cal F}$. The numerical techniques we use are explained in some detail in Sections~\ref{sec:Deltanum} and \ref{sec:Fnum}, respectively, and we hope they may also find some use in other similar problems.

The structure of the paper is as follows. In Section~\ref{sec:N2summary} we summarize some relevant facts about correlation functions in the $\mathbf{E}$ theory and their calculations by supersymmetric localization techniques. In Section~\ref{sec:Deltakinteq}, we develop a method involving integral equations to calculate $\Delta_k(\lambda)$. We solve the resulting integral equation numerically to calculate $\Delta_k(\lambda)$ accurately for many values of $\lambda$ and we use these results to conjecture analytic expressions for the large $\lambda$ behaviour of $\Delta_k(\lambda)$. In Section~\ref{sec:FreeEnergy}, we rewrite the free energy $\cal{F}$ as a Fredholm determinant and explain how to calculate it numerically for many values of $\lambda$. We again use these numerical results to conjecture the large $\lambda$ behaviour of $\cal{F}$ and therefore for $\delta_k(\lambda)$ and $\Delta q(\lambda)$. In Section~\ref{sec:Threeptfc}, we show how to use these results to determine the large $\lambda$ behaviour of three point extremal correlators. We conclude in Section~\ref{sec:Discussion} with a short discussion on some open questions for future research. The four appendices contain some details on our numerical algorithm as well as an illustrative example of an analytic approach to the calculation of $\Delta_k$ which we believe is wrong for subtle reasons.

\textit{Note added:} While we were finalizing the first version of this manuscript \cite{Beccaria:2022ypy} appeared which has overlap with some of the results presented in Section~\ref{sec:Deltakinteq} and Section~\ref{sec:FreeEnergy} below. The results in \cite{Beccaria:2022ypy} were derived by different methods but agree with ours. We have added some comments in the second version of this manuscript in order to facilitate comparison with the results of \cite{Beccaria:2022ypy}.

\section{$\cN = 2$ conformal field theories and matrix models}
\label{sec:N2summary}

In this section, we briefly recall some background material on supersymmetric localization and matrix model results for correlation functions in 4d $\mathcal{N}=2$ SCFTs. More details on this material can be found in \cite{Billo1, Billo:2022xas,Billo:2019st,Gerchkovitz:2016bo,Beccaria:2007ln,Baggio:2014sna}.

An interesting set of objects in $\cN = 2$ SCFTs are single trace chiral primary operators built out of the complex scalar $\varphi$ in the vector multiplet:
		\begin{equation}\label{CPO}
			O_n(x) \equiv {\rm tr}\,\varphi^n(x)\,.
		\end{equation}
		These operators are chiral because they are annihilated by half of the supersymmetries, and they are automatically normal-ordered because of $R$-charge conservation. The anti-chiral primary operators are denoted by $\bar{O}_n(x)$ and are constructed from $\bar{\varphi}(x)$ in the same way. The properties of the OPE for $4d\ \cN = 2$ theories leads to the chiral-ring relation:
		\begin{equation}\label{ChiralRing}
			O_m(x) O_n(0) = O_{m,n}(x) + \cdots,\quad O_{m,n}(x) \equiv {\rm tr}\, \varphi^m(x) {\rm tr}\, \varphi^n(x),
		\end{equation}
		where ``$\cdots$'' denotes $\bar{Q}$-exact terms. In this paper, we are interested in the two-point  and three-point functions involving chiral and anti-chiral operators. These correlators are constrained by the chiral-ring relation and R-symmetry selection rules and take the following form
		\begin{equation}\label{FTCorrelators}
			\langle O_m(0) \bar{O}_n(x)\rangle = \frac{G_m(N, \lambda) \delta _{m,n}}{(4\pi^2 x^2)^{2m}},\qquad 	\langle O_m(0)O_n(x) \bar{O}_p(y)\rangle = \frac{G_{m,n}(N, \lambda)  \delta _{m+n,p }}{(4\pi^2 x^2)^{m}(4\pi^2 y^2)^{n}  }.
		\end{equation}
	One can use supersymmetric localization techniques\footnote{There are two localization schemes in the literature; in this paper we use the full Lie algebra localization \cite{Billo1, Billo2,Billo:2022xas, Beccaria:2007ln, Billo:2019st}.} to study ${\cal N} = 2$ SCFTs, see \cite{Pestun:2016zxk} for a review. For example, in the planar limit, the partition function on $S^4$ is given by the following matrix integral:
		\begin{equation}\label{PartitionFunction}
		\cZ_{S^4} = \int da\,  e^{- {\rm tr}\, a^2 - S_{\rm int}(a) }\ .
	\end{equation}
In this integral, $a = a^bT_b$ with $T_b$ the generators of the gauge algebra, and the measure is given by
\begin{equation}
		da = \prod_{b=1}^{N^2-1} \frac{da^b}{\sqrt{2\pi}}\,.
\end{equation}
One can show that $S_{\rm int}(a)=0$ for the ${\cal N} = 4$ SYM theory and the matrix model is therefore Gaussian in that case. There is a closed form expression for the interaction term $S_{\rm int}(a)$ in general. The general form of this expression is not presented here, but can be found for the {\bf E} theory in equation~\eqref{EtheorySint} below.

For 4d $\mathcal{N}=2$ $\mathfrak{su}(N)$ gauge theories with $N_F$ hypermultiplets in the fundamental representation, $N_A$ hypermultiplets in the anti-symmetric representation and $N_S$ in the symmetric representation, the gauge coupling admits corrections only at 1-loop order, and the coefficient of its $\beta$-function is
	\begin{equation}
		\beta_0 = 2N - N_F - (N+2)N_S - (N-2)N_A \,.
	\end{equation}
	There exist five families of theories with vanishing $\beta_0$ \cite{Koh:1984}, which are called {\bf ABCDE} theories respectively. Their matter content is listed below:
	$$
	\begin{array}{|c||c|c|c|c|c|}
		\hline
		{\rm theory}  & {\rm\bf A} & {\rm\bf B} & {\rm\bf C} & {\rm\bf D} & {\rm\bf E} \\
		\hline 
		\hline
		N_F & 2N & N-2 & N+2 & 4 & 0\\
		N_S & 0 & 1 & 0 & 0 & 1\\
		N_A & 0 & 0 & 1 & 2 & 1\\
		\hline
	\end{array}
	$$
	In this paper we will only study the {\bf E} theory. One can calculate that for this theory, the interaction term in the matrix model~\eqref{PartitionFunction} is given by the following expression
	\begin{equation}\label{EtheorySint}
		S_{\rm int}^{\rm \bf E} (a)= -4 \sum_{l,m = 1}^\infty \left(- \frac{g^2 }{8\pi^2 }  \right)^{l+m+1} \frac{(2l+2m+1)!}{(2l+1)!(2m+1)!}  \zeta(2l+2m+1) {\rm tr}\, a^{2l+1}\, {\rm tr}\, a^{2m+1},
	\end{equation}
where $\zeta(s)$ is the Riemann zeta function. 
	Expectation values in the matrix model are given by the usual expression
		\begin{equation}\label{EMM}
				\langle f(a) \rangle = \frac{\int da\, f(a)\, e^{-{\rm tr}\, a^2 - S_{\rm int}(a)} }{ \int da\, e^{-{\rm tr}\, a^2 - S_{\rm int}(a)}  } \,.
			\end{equation}
		This can also be written as
		\begin{equation}
			\langle f(a) \rangle = \frac{\left\langle f(a) e^{-S_{\rm int}} \right\rangle_{(0)} }{  \left\langle e^{-S_{\rm int}} \right\rangle_{(0)} },
		\end{equation}
		with the expectation value in the Gaussian model defined as
		\begin{equation}
			\langle f(a) \rangle_{(0)} =\int da e^{-{\rm tr}\, a^2} f(a) \,.
		\end{equation}
	One can also use the matrix model to calculate the correlation functions~\eqref{FTCorrelators}. Namely, in the planar limit, one finds 
	\begin{equation}\label{Def3ptFunctions}
		G_n =  \langle O_n(a) O_n(a) \rangle\,, \quad\text{and}\quad G_{m,n} =  \langle O_m(a) O_n(a) O_{m+n}(a) \rangle\,.
	\end{equation}
	The expectation values on the right hand sides are given by~\eqref{EMM}. The operators $O_n(a)$ are defined as \cite{Billo:2022xas, Gerchkovitz:2016bo}
	\begin{equation}\label{O2Omega}
	O_n(a) = \Omega_n(a) - \sum_{m < n} C_{n,m}O_m(a)\,,
\end{equation} 
	with $\Omega_n(a) = {\rm tr}\, a^{n}$ and the mixing coefficients $C_{n,m}$ given by
	\begin{equation}\label{mixingCmn}
		C_{n,m} =\frac{\langle \Omega_n(a) \Omega_m(a) \rangle }{\langle \Omega_m(a) \Omega_m(a) \rangle }\,.
	\end{equation} 
Equation~\eqref{O2Omega} can be viewed as resulting from applying Gram-Schmidt orthogonalization. 
	
From now on, we will denote the quantities related to the zero coupling limit with a suffix $^{(0)}$, which is identical to $\cN = 4$ YM. Then we define the normalized normal-ordered operators in the matrix model, focusing on the twisted sector:
	\begin{equation}\label{Defomega}
	\omega_k(a) \equiv O_{2k+1}^{(0)}(a) \Big/ \sqrt{G_{2k+1}^{(0)}},\quad \langle \omega_k(a) \omega_l(a) \rangle_{(0)} = \delta_{kl} ,
	\end{equation}
	where $G_{2k+1}^{(0)} = (2k+1)(N/2)^{2k+1}$ in the large $N$ limit.	As is pointed out in \cite{Beccaria:2007ln}, the correlators of $\omega_k(a)$ can be evaluated using Wick's theorem. We can regard the matrix operators $\omega_k(a)$ as a set of normally distributed real variables $\omega_k$ and write
	\begin{equation}\label{FreePFInomega}
		\langle \omega_{k_1}(a)  \omega_{k_2}(a) \cdots  \omega_{k_n}(a) \rangle = \int [\cD \omega]\omega_{k_1} \omega_{k_2}\cdots \omega_{k_n} e^{-\frac{1}{2}\mathbf{\omega}^T\mathbf{\omega}} ,\quad [\cD\omega]\equiv  \prod_{i=1}^\infty \frac{d\omega_i}{\sqrt{2\pi}}\,,
	\end{equation}
	where we denote $\omega$ as an infinite vector whose components are $\omega_k$. It was pointed out in \cite{Beccaria:2007ln} that one can re-express $S_{\rm int}$ for {\bf E} theory \eqref{EtheorySint} in terms of $\omega$ as
	\begin{equation}
		S^{\rm \bf E}_{\rm int} = - \frac{1}{2} \omega^T \mathsf{X}\,  \omega\,,
	\end{equation}
	where the infinite matrix $\mathsf{X}$ is given in \eqref{Xx}.
	
	We will study several observables using this formalism. As an example, the partition function of the matrix model \eqref{PartitionFunction}, ignoring the normalization factor, is given by:\footnote{The fact that $S_{\rm int}$ of {\bf E} theory only contains odd double-traces is important here and is the main reason we restrict ourselves to the {\bf E} theory. }
	\begin{equation}
		\cZ = \left\langle e^{-S_{\rm int}} \right\rangle_{(0)} =  \left\langle e^{\frac{1}{2} \omega^T \mathsf{X} \omega } \,.\right\rangle_{(0)} 
	\end{equation}
	Using \eqref{FreePFInomega}, one obtains:
	\begin{equation}
	\cZ = \int [\cD\omega] e^{-\frac{1}{2}\omega^T (\mathbf{1} - \mathsf{X}) \omega} = {\rm det}^{-1/2}\left(\mathbf{1} - \mathsf{X} \right) \,.
	\end{equation}
	The corresponding free energy, which is actually the free energy of the ``difference theory'', is given by \cite{Tseytlin}
	\begin{equation}
		{\cal F} \equiv F^{\rm E-theory} - F^{\cN=4} = -\log\cZ = \frac{1}{2} \log \det ( 1 - \mathsf{X})\,.
		\label{FreeEnergyDifference}
	\end{equation} 

\section{Twisted correlators}\label{sec:Deltakinteq}

		Following the last section, the expectation value of any operator $f(\omega(a))$ containing only odd traces can be written in terms of the free model quantities as
		\begin{equation}
		\langle f(\omega(a)) \rangle = \frac{1}{\cZ } \int [\cD\omega] f(\omega) e^{-\frac{1}{2} \omega^T (\mathbf{1} - \mathsf{X}) \omega} \,.
		\end{equation}
		Thus the propagators of $\omega_k(a)$ in the interacting theory are given by 
		\begin{equation}\label{2ptFunctionomega}
		\langle \omega_k (a) \omega_l(a) \rangle =  \Big(\frac{1}{1 - \mathsf{X}}\Big)_{k,l } \equiv	 \Dx_{k,l} \,.
		\end{equation}
		Multiple correlators can be obtained by Wick contraction with propagator $\mathsf{D}$. 
		
		What we are really interested in are the operators $O_n(a)$. Thus we define the analogue of $\omega_k(a)$, which are by definition diagonal and suitably normalized:
		\begin{equation}
			\widehat{\omega}_k(a) \equiv \frac{O_{2k+1}(a) } {\sqrt{G_{2k+1}^{(0)}} },\quad \langle \widehat{\omega}_k(a) \widehat{\omega}_l(a) \rangle = \frac{\langle {\rm tr}\, \varphi^{2k+1}\, {\rm tr}\, \bar{\varphi}^{2l+1} \rangle }{ \langle {\rm tr}\, \varphi^{2k+1}\, {\rm tr}\, \bar{\varphi}^{2l+1} \rangle_{(0)}  } = \delta_{kl} \left( 1 + \Delta_k(\lambda) + O(1/N)^2\right)\,.
		\end{equation}
		Comparing with \eqref{Defomega}, we find that $\Delta_k(0) = 0$. As pointed out in \cite{Billo1}, from the Gram-Schmidt procedure, one finds from \eqref{2ptFunctionomega} that:\footnote{Here we use $A_{(k)}$ to denote the upper left $k \times k$ block of the matrix $A$. }
	\begin{align}\label{defDelta}
		1 + \Delta_k(\lambda) = 
		\frac{\det \Dx_{(k)}}{\det \Dx_{(k-1)}}\,.
	\end{align}
		In the remainder of this section, we will introduce a novel method to evaluate the matrix elements $\mathsf{D}_{kl}$ and the quantity $\Delta_k$ both numerically and analytically.

\subsection{Fredholm integral equations}
\label{sec:FredhomeIntegralEqn}

To calculate  $\Delta_k(\lambda)$ numerically one could truncate the matrix \eqref{Xx} to size $M\times M$ with $M$ large. The calculation of \eqref{defDelta} then only involves linear algebra with finite matrices. However, for large $M$, there are many integrals to calculate numerically which is slow. We therefore proceed as follows.
If $V$ is an $n \times m$ matrix, it is easy to check that
\begin{equation}\label{eq:eq1}
	(1 + V V^T)^{-1} = 1 - V(1 + V^T V)^{-1} V^{T}\,.
\end{equation}
Notice that the inverse on the left is of an $n\times n$ matrix, whereas the inverse on the right is of an $m \times m$ matrix.
This can also be written as
\begin{equation}\label{eq:eq2}
	(1 + V V^T)^{-1} = 1 - V Z\,,
\end{equation}
where $Z$ is the solution of the equation 
$ (1 + V^T V) Z = V^{T}$.

A limiting procedure for $m \to \infty$ then gives the following result. If $V_k(t)$ are functions with $k = 1 , \ldots, n$ and $t \in [ 0, + \infty[ $ and if we define a matrix $V V^T$ as
\begin{equation}
	( V V^T)_{kl} = \int_0^{+\infty}\!\!\!\! dt\ V_k(t)\ V_l(t)\,, 
	\label{VVT}
\end{equation}
then
\begin{equation}\label{eq:eq3}
	(1 + V V^T)^{-1}_{kl} = \delta_{kl} -  \int_0^{+\infty}\!\!\!\! dt\ V_k(t)\ Z_l(t)\,,
\end{equation}
where $Z_k(t)$ is the solution of the integral equation
\begin{equation}\label{eq:eq4}
	Z_k(t) +  \int_0^{+\infty}\!\!\!\! ds\  M(t,s) Z_k(s)= V_k(t)\,,
\end{equation}
with $ M(t,s) = \sum_{k=1}^n V_k(t) V_k(s)$. The matrix $\mathsf{X}$ defined in \eqref{Xx} has the form \eqref{VVT}, thus its inverse matrix $\mathsf{D}$ can be written using \eqref{eq:eq3} as
\begin{equation}\label{eq:eq6}
	\Dx_{kl} = \delta_{kl} - (-1)^{k+l} \sqrt{2k+1} \sqrt{2l+1}  
	\int_0^{+\infty}\!\!\!\! dt\ v(t) J_{2 k+1} (t) \Psi_{2 l+1} (t)\,,
\end{equation}
where  $\Psi_{2 l+1} (t)$ is the solution of the integral equation
\begin{equation}
		\Psi_{2 k+1} (t) + \int_0^{+\infty}\!\!\!\! ds\ K(t,s) \Psi_{2 k+1} (s) = v(t) J_{2 k+1} (t)\,, \label{eq:eq7}
\end{equation}
with
\begin{equation}
		\qquad K(t,s) = v(t) K_B(t,s) v(s)\,,\qquad\text{and}\qquad 	v(t)^2 = \dfrac{16 \pi}{\sqrt{\lambda}}\ W\left(\dfrac{2 \pi t}{\sqrt{\lambda}}\right)\,.\nonumber
\end{equation}
	Here $K_B(t,s)$ is the Bessel kernel \cite{TracyWidom}, which can be evaluated in closed form as \cite{Billo1}\footnote{A short proof of this identity can be found in \cite{Tygert}.}: 
\begin{equation}\label{eq:kernelK2}
	K_B(t,s) \equiv \sum_{k=1}^{+\infty} (2k+1)J_{2 k+1}(t) J_{2 k +1}(s) =  - \frac{1}{2} \frac{t s}{t^2 - s^2} \left(t J_1(t) J_2(s) - s J_2(t)J_1(s)\right)\,.
\end{equation}
All in all, we have converted the original definition \eqref{2ptFunctionomega} of $\Dx_{kl}$, which uses the inverse of an infinite matrix,
to an expression which uses instead the solution of an integral equation. In Section~\ref{sec:smallDeltaAN} we will perform a check of  the equations derived above by comparing against the expansion in small $\lambda$. 
We are not able to solve the integral equation \eqref{eq:eq7} analytically, therefore we resort to a numerical method which we discuss in Section~\ref{sec:Deltanum}.

\subsection{Analytical comparison against small $\lambda$ expansion}
\label{sec:smallDeltaAN}

In this section we calculate the small $\lambda$ expansion of $\Dx_{k,l}$ using equations \eqref{eq:eq6} and \eqref{eq:eq7}. For convenience, in this section, we use the notation $\mu = \dfrac{\sqrt{\lambda}}{2 \pi}$. We also change variables $t / \mu = x$ and  
$s/ \mu = y$ and define $ \chi_{2 k+1}(x) = \Psi_{2 k+1}(\mu x)$.
Equations \eqref{eq:eq6} and \eqref{eq:eq7} then become
\begin{equation}\label{small:D}
	\Dx_{kl} = \delta_{kl} - (-1)^{k+l} \sqrt{2k+1} \sqrt{2l+1}\ ( 8 \mu)^{1/2}  
	\int_0^{+\infty}\!\!\!\! dx\ W(x)^{1/2} J_{2 k+1} (\mu x) \chi_{2 l+1} (x)\,,
\end{equation}
and
\begin{equation}\label{small:inteq}
	\chi_{2 k+1} (x) + 8 \int_0^{+\infty}\!\!\!\! dy\ W(x)^{1/2}\ K_B(\mu x, \mu y) W(y)^{1/2}\chi_{2 k+1} (y) 
	= \left(\frac{8}{\mu} \right)^{1/2} W(x)^{1/2}\ J_{2 k+1} (\mu x) \,.
\end{equation}
Using the Taylor series expansion of Bessel functions, equation \eqref{eq:kernelK2} leads to the following small $\lambda$ (or equivalently small $\mu$) expansion
$$
K_B(\mu x, \mu y) = \frac{1}{768} x^3 y^3 \mu^6 - \frac{1}{12288} ( x^5 y^3 + x^3 y^5  ) \mu^8 +O(\mu)^{10}\,.$$
Therefore, up to this order, the kernel of the integral operator in \eqref{small:inteq} is degenerate, and the integral equation 
can be solved exactly.\footnote{See Appendix~\ref{sec:degkernels} for more details on how to implement this and for the notation we use.}
We can thus write
\begin{equation}
K(x,y) = \sum_{i=1}^3 a_i(x) b_i(y) + O(\mu)^{10}\,,
\end{equation}
with
\begin{align*}
	a_1(x) & = \frac{8}{768} x^3 W(x)^{1/2}\,, & b_1(y) & = y^3 W(y)^{1/2}\mu^6\,, \\
	a_2(x) & = - \frac{8}{12288}  x^5 W(x)^{1/2}\,, & b_2(y) & = y^3 W(y)^{1/2}\mu^8\,, \\
	a_3(x) & = - \frac{8}{12288}  x^3 W(x)^{1/2}\,, & b_3(y) & = y^5 W(y)^{1/2}\mu^8\,.
\end{align*}
The $3 \times 3 $ matrix
\begin{equation}
	A_{ki} = \int_0^{+\infty}\!\!\!\! dx\ b_k(x) a_i(x)\,,
\end{equation}
can be calculated analytically by employing the Mellin transform
\begin{equation}\label{small:Mellin}
	\int_0^{+\infty}\!\! \frac{dx}{x} x^p\ W(x) = \Gamma(p-1) \zeta(p-2)\,,
\end{equation}
to find
\begin{equation}\label{small:A}
	A = 
	\left(
	\begin{array}{ccc}
		\dfrac{5  }{4} \zeta (5)\mu ^6& -\dfrac{105}{32} \zeta (7) \mu ^6 & -\dfrac{5}{64} \zeta (5) \mu ^6 \\[4mm]
		\dfrac{5 }{4} \zeta (5) \mu ^8 & -\dfrac{105}{32}  \zeta (7) \mu ^8 & -\dfrac{5}{64} \zeta (5) \mu ^8 \\[4mm]
		\dfrac{105 }{2} \zeta (7) \mu ^8 & -\dfrac{945}{4}  \zeta (9) \mu ^8 & -\dfrac{105}{32} \zeta (7) \mu ^8 \\
	\end{array}
	\right)+O(\mu)^{10}\,.
\end{equation}
The inverse is given by
$$
(1 + A)^{-1} =  
\left(
\begin{array}{ccc}
	1-\dfrac{5 }{4} \zeta (5)\mu ^6  & \dfrac{105 }{32}\zeta (7)\mu ^6  & \dfrac{5}{64} \zeta (5)\mu ^6  \\[4mm]
	-\dfrac{5 }{4}\zeta (5)\mu ^8  & 1+\dfrac{105 }{32}\zeta (7)\mu ^8  & \dfrac{5 }{64}\zeta (5)\mu ^8  \\[4mm]
	-\dfrac{105 }{2}\zeta (7)\mu ^8  & \dfrac{945 }{4}\zeta (9)\mu ^8  & 1+\dfrac{105 }{32}\zeta (7)\mu ^8
\end{array}
\right)+O(\mu)^{10}\,.
$$
Equation \eqref{eq:deg8} then gives
\begin{multline*}
\chi_{2 k +1}(x) =  \left(\frac{8}{\mu} \right)^{1/2}W(x)^{1/2} J_{2 k+1} (\mu x) \\
- \sum_{i,j=1}^3 \int_0^{+\infty}\!\!\!\! dy\ a_i(x) ( 1 + A)^{-1}_{ij} b_j(y)  \left(\frac{8}{\mu} \right)^{1/2} W(y)^{1/2} J_{2 k+1} (\mu y)\,.
\end{multline*}
 
One can now expand the Bessel functions again as a series in small $\mu$ and calculate the integral with \eqref{small:Mellin}. This gives an expansion for $\chi_{2 k +1}(x)$ at small $\mu$. The result of this procedure is fairly complicated and we do not present it explicitly here. Finally, inserting $\chi_{2 k +1}(x)$ into equation \eqref{small:D} gives $\Dx_{kl}$ as a series in $\mu$:
\begin{align*}
	D_{11} & = 1-\frac{5 }{4}\zeta (5)\mu ^6 +\frac{105}{16} \zeta (7)\mu ^8 -\frac{1701}{64} \zeta (9)\mu ^{10}
	+ \left(\frac{25 }{16}\zeta (5)^2+\frac{12705}{128} \zeta (11)\right)\mu ^{12}+O\left(\mu\right)^{13}\,,\\[4mm]
	D_{12}&=D_{21} = \frac{7}{32} \sqrt{15}\ \zeta (7)\mu ^8 -\frac{105}{64} \sqrt{15}\ \zeta (9)\mu ^{10} +\frac{1089}{128} \sqrt{15}\  \zeta (11) \mu ^{12}+O\left(\mu\right)^{13}\,,\\[4mm]
	D_{22}&= 1-\frac{63}{64} \zeta (9)\mu ^{10} +\frac{1155 }{128}\zeta (11)\mu ^{12} +O\left(\mu\right)^{13}\,.
\end{align*}
This then finally leads to 
\begin{align*}
	\Delta_1(\lambda) & = -\frac{5 }{4}\zeta (5)\mu ^6 +\frac{105}{16} \zeta (7)\mu ^8 -\frac{1701}{64} \zeta (9)\mu ^{10}
	+ \left(\frac{25 }{16}\zeta (5)^2+\frac{12705}{128} \zeta (11)\right)\mu ^{12}+O\left(\mu\right)^{13}\,,\\[4mm]
	\Delta_2(\lambda) & =-\frac{63}{64} \zeta (9)\mu ^{10} +\frac{1155 }{128}\zeta (11)\mu ^{12} -\frac{27885 }{512} \zeta (13)\mu ^{14}+O\left(\mu\right)^{15}\,.
\end{align*}
This agrees with equations (3.37) and (3.38) in \cite{Billo1}. Since the results in \cite{Billo1} were obtained by a different method we view this as a consistency check of our approach.

\subsection{Numerical method for the calculation of $\mathsf{D}_{kl}$ and $\Delta_k(\lambda)$}\label{sec:Deltanum}

To solve the integral equation \eqref{eq:eq7} we employ numerics and use the Nystr\"om method. 
This method is well known and is based on discretizing the integral appearing in the integral equation in the schematic form
\begin{equation}\label{eq:eq11}
	\int_0^{+\infty}\!\!\!\! dt\ f(t) \approx \sum_{a=1}^m w_a f(t_a)\quad\text{with $m$ large}\,.
\end{equation}
Here, $w_a \ge 0$ are weights and $t_a$ are the discretization points. The upper left $p \times p$ block of the matrix $\Dx$ is then equal to
\begin{equation}\label{eq:eq12}
	\Dx_{(p)} = \mathbf{1}_{ p \times p} - \mathbb{V} \left( \mathbf{1}_{m \times m} + \mathbb{K}\right)^{-1} \mathbb{V}^T\,,
\end{equation}
where $\mathbb{V}$ is a $p \times m$ matrix and $\mathbb{K}$ is an $m \times m$ matrix given by
\begin{equation}
\begin{split}
	\mathbb{V}_{ka} &= \sqrt{w_a}\ (-1)^k \sqrt{2 k+1}\ J_{2 k+1}(t_a)\ v(t_a)\,,\\
\mathbb{K}_{ab} &= \sqrt{w_a}\ K(t_a, t_b)\ \sqrt{w_b}\,.	
\end{split}
\end{equation}
If $m \to \infty$, \eqref{eq:eq12} should converge to the correct result. 
All in all, what we have done is instead of truncating the original matrix \eqref{Xx} to size $M \times M$ and treating the integrals over the Bessel functions exactly, we have taken $M \to \infty$ and discretized the integrals. 

\begin{figure}[ht]
	\centering
	\includegraphics[scale = .9]{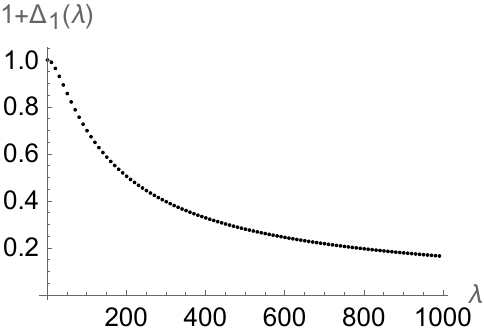} \includegraphics[scale = .9]{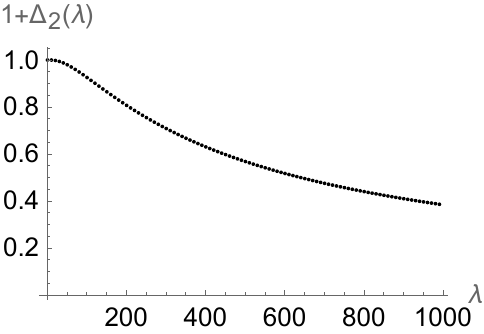}
	\caption{ $1 + \Delta_k(\lambda)$ as function of $\lambda$ with $k=1$ (left) and $k=2$ (right).}\label{fig1}
\end{figure}

\begin{figure}[H]
	\centering
	\includegraphics[scale = .9]{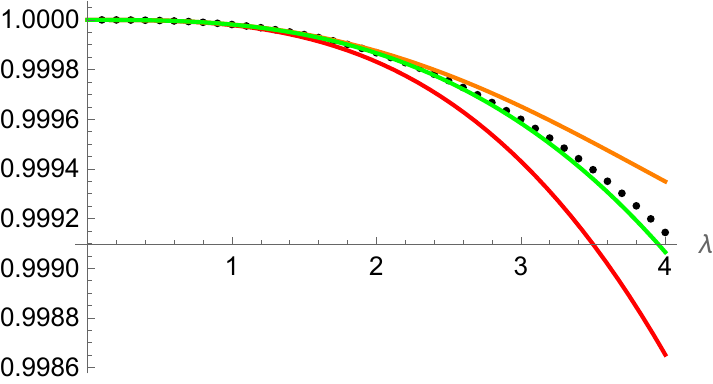}
	\caption{Black dots: numerical results for $1+\Delta_1(\lambda)$.	Red line: $ 1-\frac{5\,\zeta (5)}{256\pi^6}\lambda^3$;
	Orange line: $ 1-\frac{5\,\zeta (5)}{256\pi^6}\lambda^3+\frac{105\, \zeta (7)}{4096\pi^8}\lambda^4$;
	Green line: $1-\frac{5\,\zeta (5)}{256\pi^6}\lambda^3+\frac{105\, \zeta (7)}{4096\pi^8}\lambda^4-\frac{1701\, \zeta (9)}{65536\pi^{10}} \lambda^5$.}\label{Figure0630a}
\end{figure}

There are many discretization schemes (also known as quadrature rules) that one can use in~\eqref{eq:eq11}. Some of these are discussed in Appendix \ref{app:discretisation}. We chose the Fej\'er type 1 quadrature rule. In this rule,  we also have to truncate the integral
$$\int_0^{+\infty}\!\!\! dt\ \ f(t) \approx \int_0^L \!\!\! dt\ \ f(t) \approx \sum_{a=1}^m w_a f(x_a)\,.$$
So we have to take $L$ and $m$ both large to get accurate results. More details on Fej\'er type 1 and the values of $L$ and $m$ can be found in
Appendix \ref{app:discretisation}.
As an illustration, the two graphs in Figure \ref{fig1} were made using equation~\eqref{eq:eq12} with appropriate settings for $L$ and $m$.
As a check on the  accuracy of the numerical algorithm we can compare our numerical result against the analytic calculations of the small $\lambda$ expansion of $\Delta_k(\lambda)$ discussed in \cite{Billo1}. The first few terms in the small $\lambda$ expansion for $\Delta_1$ read 
\begin{equation}\label{Delta1small}
	\Delta_1(\lambda)  =-\frac{5\,\zeta (5)}{256\pi^6}\lambda^3+\frac{105\, \zeta (7)}{4096\pi^8}\lambda^4-\frac{1701\, \zeta (9)}{65536\pi^{10}} \lambda^5+\cdots\,.
\end{equation}
In Figure~\ref{Figure0630a} we compare this analytic result against our numerics. It is clear that there is very good agreement between the two calculations for small values of $\lambda$.\footnote{Since the numerical coefficients multiplying the powers of $\lambda$ in \eqref{Delta1small} are small ($10^{-5}$ or smaller) we can treat order 1 values for $\lambda$ as ``small''.} 

As discussed in \cite{Beccaria:2007ln}, the radius of convergence of the series for $\Delta_1$ is $\lambda_c = \pi^2$ because of the branch point located at $\lambda = -\pi^2$, see also \cite{Russo:2013sba}.

In Table~\ref{tab:smallDelta1} we provide additional numerical evidence for the agreement between the two methods. 
\begin{table}[H]
	\centering
	$$
	\begin{array}{|l|lllll|}
		\hline
		n \backslash \lambda & 1 & 2 & 3 & 4 & 20 \\
		\hline
		1 & -0.0000210659 & -0.000168527 & -0.000568778 & -0.00134822 & -0.168527 \\
		2 & -0.0000183417 & -0.000124939 & -0.000348117 & -0.000650816 & 0.267347 \\
		3 & -0.0000186194 & -0.000133826 & -0.000415601 & -0.000935195 & -0.621336\\
		4 & -0.0000185927 & -0.000132119 & -0.000396155 & -0.000825934 & 1.08587 \\
		5 & -0.0000185952 & -0.000132442 & -0.000401667 & -0.000867224 & -2.1399 \\
		\hline
		\text{numerics} & -0.000018595 & -0.000132391 & -0.000400453 & -0.000855928 &
		-0.0326525 \\
		\hline
	\end{array}
	$$
	\caption{Comparison of the small $\lambda$ expansion against numerical calculations of $\Delta_1(\lambda)$ for different values of $\lambda$. The last row gives the value which is calculated with the numerical method explained in Section~\ref{sec:Deltanum}. The first row with $n=1$ gives the result using 1 term in the expansion~\eqref{Delta1small}, the second row with $n=2$ gives the result using 2 terms in the expansion~\eqref{Delta1small} and so on. It is clear that for small $\lambda < \lambda_c $ ($\lambda = 1,2,3,4$ in the table) the series~\eqref{Delta1small} converges to the numerical result as $n$ is increased. For large $\lambda > \lambda_c $ ($\lambda = 20$ in the table), the series~\eqref{Delta1small} does not converge, but the numerical method of Section~\ref{sec:Deltanum} still works.}\label{tab:smallDelta1}
\end{table}

We can proceed similarly and test our method for $k=2$, see Table~\ref{tab:smallDelta2}. To do this we use the result for the small $\lambda$ expansion of $\Delta_2$ from \cite{Billo1} 
\begin{equation}\label{Delta2small}
	\Delta_2(\lambda)  =-\frac{63\,\zeta (9)}{65536\pi^{10}}\lambda^5+\frac{1155\, \zeta (11)}{524288\pi^{12}}\lambda^6-\frac{27885\, \zeta (13)}{8388608\pi^{14}} \lambda^7+\cdots \,.
\end{equation}

\begin{table}[H]
	\centering
	$$
	{\small \begin{array}{|l|lllll|}
		\hline
		n \backslash \lambda & 1 & 2 & 3 & 4 & 20 \\
		\hline
		1 & -1.02857\times 10^{-8} & -3.29142\times 10^{-7} & -2.49942\times 10^{-6}
		& -1.05325\times 10^{-5} & -3.29142\times 10^{-2} \\
		2 & -7.90102\times 10^{-9} & -1.76523\times 10^{-7} & -7.60998\times 10^{-7}
		& -7.64937\times 10^{-7} & 1.19705\times 10^{-1} \\
		3 & -8.26546\times 10^{-9} & -2.23172\times 10^{-7} & -1.55805\times 10^{-6}
		& -6.73606\times 10^{-6} & -3.46789\times 10^{-1} \\
		4 & -8.21887\times 10^{-9} & -2.11244\times 10^{-7} & -1.25232\times 10^{-6}
		& -3.68225\times 10^{-6} & 8.46105\times 10^{-1} \\
		5 & -8.22428\times 10^{-9} & -2.14016\times 10^{-7} & -1.35889\times 10^{-6}
		& -5.1016\times 10^{-6} & -1.92606 \\
		10 & -8.22374\times 10^{-9} & -2.13514\times 10^{-7} & -1.33223\times
		10^{-6} & -4.65977\times 10^{-6} & 7.26826\times 10^1 \\
		\hline
		\text{numerics} & -8.22374\times 10^{-9} & -2.13514\times 10^{-7} &
		-1.3323\times 10^{-6} & -4.66487\times 10^{-6} & -1.79469\times 10^{-3} \\
		\hline
	\end{array} }
	$$
	\caption{Comparison of the small $\lambda$ expansion against numerical calculations of $\Delta_2(\lambda)$ for different values of $\lambda$. The notation is the same as in Table~\ref{tab:smallDelta1} and the conclusions are similar.}\label{tab:smallDelta2}
\end{table}

\subsection{Large $\lambda$ expansion of $\mathsf{D}_{kl}$ and $\Delta_k$: a conjecture}
	
As outlined above and discussed in detail in \cite{Billo1} it is relatively easy to calculate the small $\lambda$ expansion of $\mathsf{D}_{kl}$. The large $\lambda$ expansion appears to be much more challenging. The first term was calculated in \cite{Billo1} and reads (assuming $k \le l$), 
		\begin{equation}
		\mathsf{D}_{kl} =  \frac{4\pi^2}{\lambda}  \sqrt{(2k+1)(2l+1)}\ k(k+1)+O(1/\lambda^{3/2})\,.
		\end{equation}
The matrix elements for $k \ge l$ follow from the symmetry $\mathsf{D}_{kl} = \mathsf{D}_{lk}$. We are not able to rigorously calculate the subleading terms in the expansion analytically. However, based on numerical investigations, we make the following conjecture for $\mathsf{D}_{kl}$ with $k \le l$,
		\begin{equation}\label{Dkl}
	\begin{aligned}
		\mathsf{D}_{kl}  = 4\pi^2 &\sqrt{(2k+1)(2l+1)} \frac{k(k+1)}{\lambda} \left\{1 - 8l(l+1) \frac{\log 2}{\lambda^{1/2}} \right.\\
		& 	\left. +\frac{2 \pi^2 }{9 \lambda} \left[ 3 l(l+1)(l-k)(l+k+1)+2(k-1)k(k+1)(k+2)  \right] \right.\\
		& \left. + \frac{16\log^22}{\lambda} l(l+1)[k(k+1)+l(l+1)-1]  \right.\\
	 	& \left. - \frac{16\pi^2\log 2}{9\lambda^{3/2}} L [K ( K -2) +L(L-2)] \right.\\ 
		& \left. - \frac{64\log^32}{9\lambda^{3/2}} L [2 L^2 + 6 K L - 7 L - 7 K + 2 K^2 +6] \right.\\
		& \left. - \frac{\zeta(3)}{3 \lambda^{3/2} } L  [3 + 4 ( 4 K^2 + 4 L^2 - 5 K - 5 L - 6 K L)] + O(\lambda)^{-2} \right\} 
	\end{aligned}
\end{equation}
We have used the notation $K = k (k+1)$ and $L = l ( l+1)$ to shorten some of the coefficients.	The expression \eqref{Dkl} was obtained as follows. First, we used the numerical method outlined in Section~\ref{sec:Deltanum} to calculate $\mathsf{D}_{kl}$ for many values of $k,l$ and many $\lambda$ ranging\footnote{For $\lambda$ greater than $e^{16}$, our algorithm converges too slowly, and the precision obtained is not good enough. } from $e^8$ to $e^{16}$. We then used this numerical data and the function {\tt LinearModelFit[]} of {\tt Mathematica} to estimate numerically the coefficients in the large $\lambda$ expansion of $\mathsf{D}_{kl}$. Finally, the precision in these fitted coefficients was sufficiently high so that we could guess\footnote{Version 1 of this paper only contained the terms up to order $\lambda^{-2}$. We have included the term of order $\lambda^{-5/2}$ in version 2. We were able to guess a closed form expression of this additional term only after the publication of \cite{Beccaria:2022ypy}. In \cite{Beccaria:2022ypy} an analytical expression for the {\it determinant} of $D_{kl}$ is obtained, this helped us to guess closed from expressions of the components $D_{kl}$ themselves. } the closed form expression \eqref{Dkl}. 
	
%
The analytical form \eqref{Dkl} for the  strong coupling behavior of $\mathsf{D}_{kl}$ is used as ingredient in the calculation of the three point function of single trace chiral/anti-chiral correlators discussed in Section \ref{sec:Threeptfc}. It can also be directly used in \eqref{defDelta} to find the quantity $\Delta_k(\lambda)$ that controls the two-point function of twisted operators. Using \eqref{defDelta} and \eqref{Dkl} we find
	\begin{equation}
		1+\Delta_k(\lambda) = \frac{8 \pi^2  k ( 2 k+1)}{\lambda} \left[1 
		-\frac{16 k\log 2}{\lambda^{1/2}}
		+\frac{32 k( 4 k -1)\log^2 2}{\lambda} +O(\lambda^{-3/2})
		\right]\,.
		\label{NikolayConjecture}
	\end{equation}
A notable fact is that the $\pi^4$ terms appearing at $O(1/\lambda^{2})$ order in $\mathsf{D}_{kl}$, see \eqref{Dkl}, cancel with each other in $\Delta_k$, leading to a relatively simple result that improves on the calculations in \cite{Billo1} by providing two more terms in the large $\lambda$ expansion. Some observations can be made based on the leading three terms. First, the powers series is in inverse powers of $\sqrt{\lambda}$, which is compatible with the string theory perturbation series in $\alpha'$ after using the identification $\lambda^{-1} \sim \alpha'^2$. Second, the coefficients of the three terms are polynomials of $k$, whose order increases by 1 with each term in the series and which exhibit a factorized form. Third, we find an increasing power of $\log 2$ in each term in the perturbative series. This factor increases the degree of transcedentality of each perturbative coefficient and perhaps suggests a renormalization of the coupling $\lambda$.

\begin{figure}[H]
	\centering
	\subfloat[][ 
	Red dots: $\log\left(1+\Delta_1(\lambda)\right)$.\\ Red line: $\log\left(\Delta_1^{(1)}(\lambda)\right)$. ]{
		\includegraphics[width=0.5\textwidth]{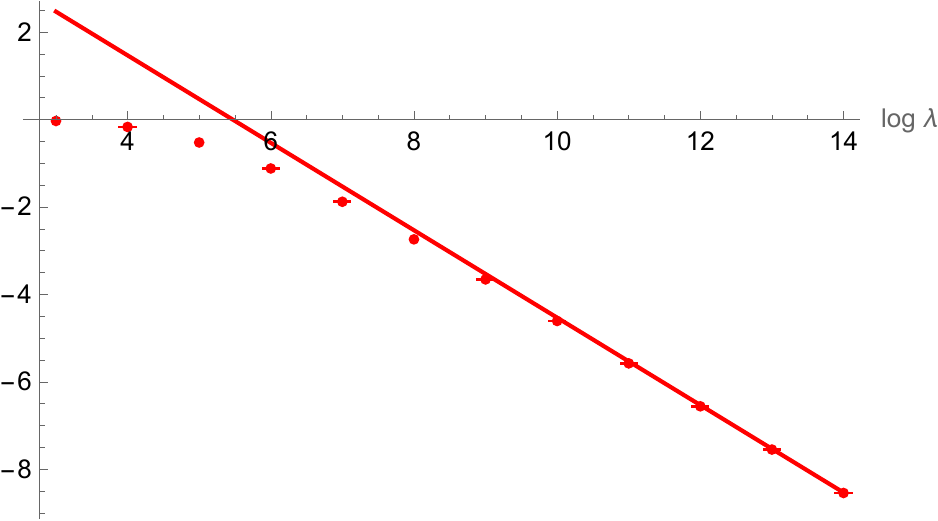}
				\label{Figure0611a}}
	\subfloat[][Orange dots: $\log\left( - (1+\Delta_1(\lambda) - \Delta_1^{(1)}(\lambda) \right) $. 
	Orange line: $\log\left( - \Delta_1^{(2)}(\lambda)\right) $. ]{
		\includegraphics[width=0.5\textwidth]{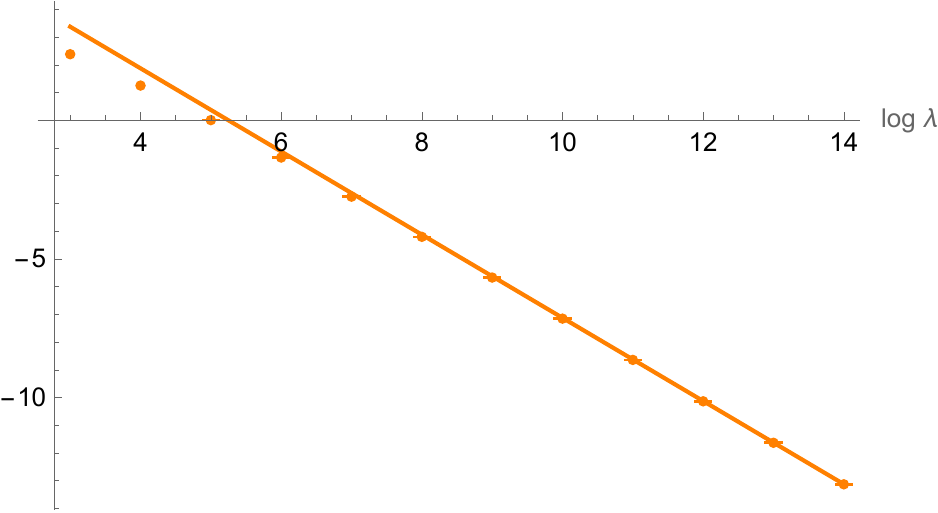}
				\label{Figure0611b}}\\
	\subfloat[][Green dots: $\log\left(1+\Delta_1(\lambda) - \Delta_1^{(1)}(\lambda) - \Delta_1^{(2)}(\lambda) \right) $. 
	Green line: $\log\left(\Delta_1^{(3)}(\lambda)\right) $.]{
		\includegraphics[width=0.5\textwidth]{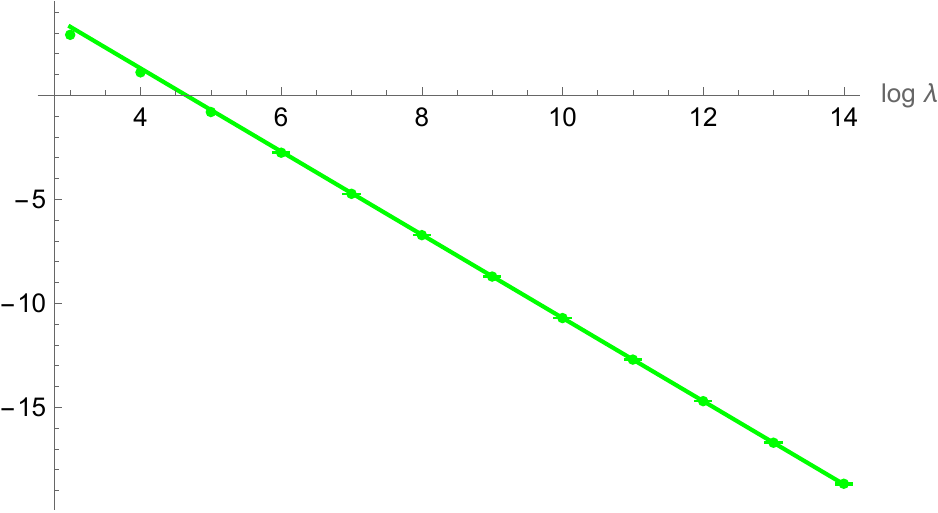}
				\label{Figure0611c}}
	\caption{A loglog plot of $1+\Delta_1(\lambda)$ and its asymptotic expansion. For large $\lambda$, the numerical data agrees very well with the asymptotic expansion. This indicates that all terms in the asymptotic expansion \eqref{NikolayConjecture} are correct.}
	\label{fig:globfig}
\end{figure}

\begin{figure}[H]
	\centering
	\includegraphics[scale = .7]{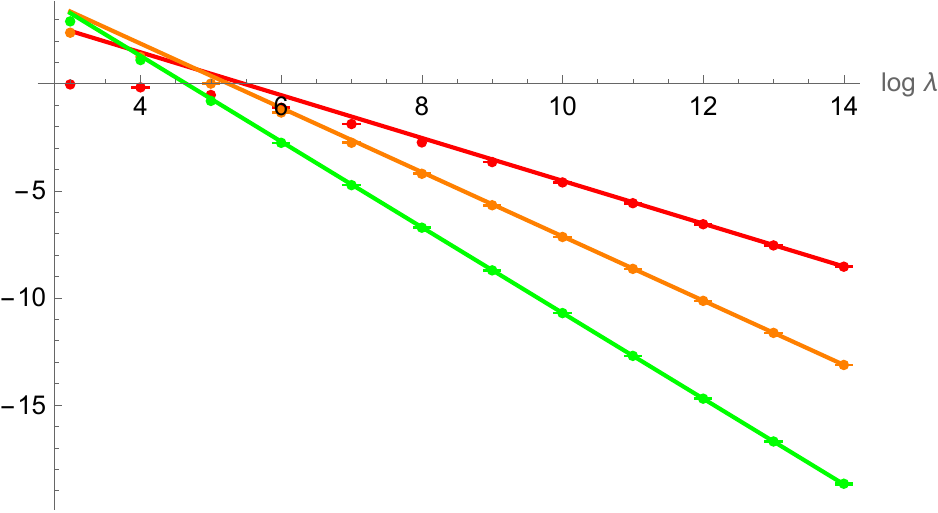}
	\caption{A combination of Figures \ref{Figure0611a}-\ref{Figure0611c}. The slope of the lines decrease if more terms in the asymptotic expansion are included. This illustrates that the coefficients in the asymptotic expansion \eqref{NikolayConjecture} are correct.  
	}\label{Figure0611d}
\end{figure}

\begin{figure}[H]
	\centering
	\includegraphics[scale = .5]{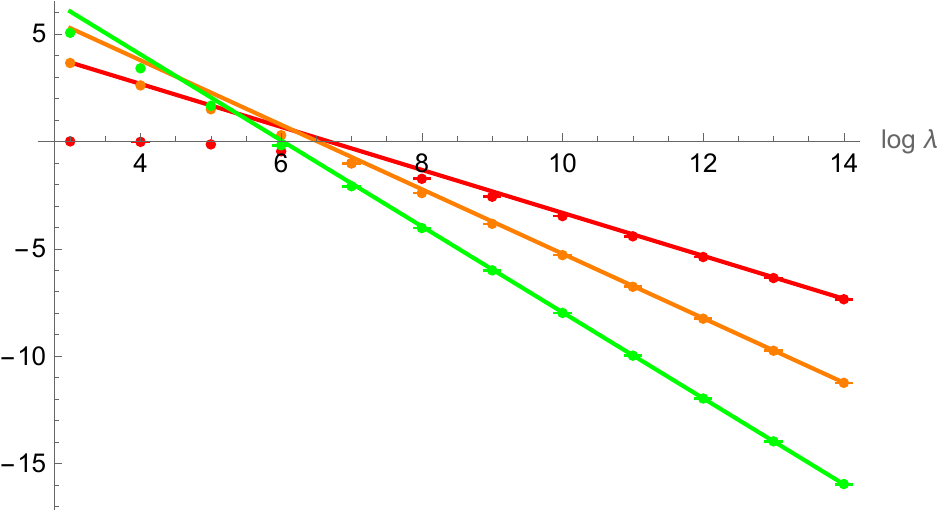}
	\includegraphics[scale = .5]{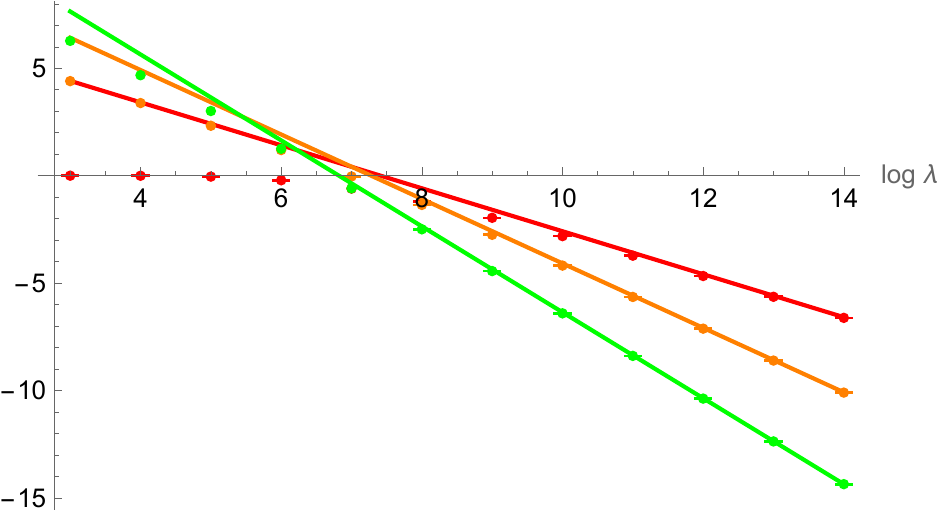}
	\caption{Similar to Figure \ref{Figure0611d}, but with $k=2$ (left) and $k=3$ (right) respectively.
	}\label{Figure0611ef}
\end{figure}

To illustrate the agreement between the analytic conjecture~\eqref{NikolayConjecture} and our numerical results we present a few plots. To facilitate the comparison we define the following three quantities that represent the three terms on the right hand side of the expansion~\eqref{NikolayConjecture}
\begin{equation}
\begin{split}
	\Delta_k^{(1)}(\lambda) & = 8 \pi^2 \left[ \frac{k ( 2 k+1)}{\lambda}
	\right]\,,\\
	\Delta_k^{(2)}(\lambda) & = 8 \pi^2 \left[-\frac{16 k^2 ( 2 k +1) \log 2}{\lambda^{3/2}}\right]\,,\\
	\Delta_k^{(3)}(\lambda) & = 8 \pi^2 \left[\frac{32 k^2 ( 2 k +1) (4 k -1)\log^2 2}{\lambda^{2}}
	\right]\,.
\end{split}
\end{equation}
In the figures below, we have added error bars on the numerical data of $1+\Delta_k(\lambda)$ with an estimation of accuracy and we have set $k=1$ in Figures~\ref{fig:globfig} and \ref{Figure0611d}. Similar results can be obtained for other values of $k$ which we illustrate for $k=2,3$ in Figure~\ref{Figure0611ef}. 

Similarly, Figure \ref{Figure0611ef} serves as evidence that the asymptotic expansion \eqref{NikolayConjecture} is valid for $k=2$ and $k=3$.

\subsection{Large $\lambda$ expansion of $\Delta_k(\lambda)$: more terms}
\label{sec:Deltamoreterms}
The matrix elements $\mathsf{D}_{kl}$ are harder to evaluate than the single-index quantity $\Delta_k$ that controls the two-point functions of physical interest.  Utilizing the same numerical method described in Section~\ref{sec:Deltanum}, we can evaluate $\Delta_k(\lambda)$ directly for $\lambda$ going up to $e^{16}$ and $k$ up to 25. A selection of the numerical data we have obtained can be found in Appendix~\ref{app:tables}. This numerical data agrees very well with the three terms in the expansion in \eqref{NikolayConjecture} as expected and allows us to find an expression for more subleading terms in the large $\lambda$ expansion of $\Delta_k(\lambda)$.
	
For future convenience, let us denote the coefficients in the large $\lambda$ expansion with $C_i$ such that $1 + \Delta_k(\lambda) = \sum_i C_i\left( \frac{1}{\lambda}\right)^i$. At $O(1/\lambda^{5/2})$, the structure of the coefficient is more or less the same as in \eqref{NikolayConjecture}, but there is a curious shift in the polynomial factor $(4k-3)$, namely we find:
\begin{equation}
	\label{C5p2}
	C_{5/2} = -8\pi^2 \cdot 32(\log^32)\cdot k^2(2k+1)(4k-1)\left[(4+4\epsilon_{5/2})k - (3-\epsilon_{5/2}) \right]\,,
\end{equation}
where $\epsilon_{5/2} \approx 0.107738$, which does not  appear to be any simple ``closed form'' number. Proceeding to higher order we find that at $O(1/\lambda^{3})$, the factor $(4k-3)$ which is shifted in $C_{5/2}$ is however unshifted, but there is a new factor $(4k-7)$ that is shifted:
\begin{equation}
	\label{C3}
	C_3 = 8\pi^2 \cdot 16 (\log^4 2) \cdot k^2  (2k+1)(4k-1)(4k-3)[(4+4\epsilon_3)k - (7-\epsilon_3)],\quad \epsilon_3 \approx -0.1381\,.
\end{equation}
	We have also observed a numerical coincidence between the two shifts. Namely, we find the following relation to be correct up to four digits:
\begin{equation}
	\epsilon_3 \approx 8 \epsilon_{5/2} - 1 \approx -0.13809664\,.
\end{equation}
The fact that we do not have closed form fully analytic expressions for $C_{5/2}$ and $C_3$ but rather have these coefficient	depend on the numerical constants $\epsilon_{5/2} $ and $\epsilon_3$ limits the accuracy of our numerical investigations due to the limited accuracy in the {\tt LinearModelFit[]} function in {\tt Mathematica}. The highest order in the large $\lambda$ expansion we could reliably evaluate is at $O(1/\lambda^{7/2})$ for which we find the coefficient 
\begin{equation}
	\label{C7p2}
	C_{7/2} = 8\pi^2 \cdot (\log^52)\cdot 2^8 k^2(2k+1)(4k-1)(4k-3)\cdot [(1+\epsilon_{7/2})k^2 - (2-2\epsilon_{7/2})]\,,
\end{equation}
where $\epsilon_{7/2} \approx 0.023$. The numerical data we used to obtain the expressions for $C_{5/2}$, $C_{3}$, and $C_{7/2}$ above is presented in Table~\ref{DataDeltakLargeLambda}.
	
	\begin{table}[h]
		\centering
		\begin{tabular}{|c||c|c|c|c|}
			\hline
			$k$ & $C_{5/2}$  & $C_{3}$ & $C_{7/2}$ & $L$ \\\hline\hline 
			1  & $-11652.$ & $-9.6\times 10^3$ & $3\times 10^4$  & 5000\\
			2 & $-70322(2\pm 2).$ & $-5.0 \times 10^{4}$ & $5 \times 10^{6}$ & 6000\\
			3 & $-6.06468 \times 10^6$& $5.83 \times 10^6$& $1.46 \times 10^8$ & 6000\\
			4 & $-2.695(60\pm 1) \times 10^7$& $5.447 \times 10^7$& $1.31 \times 10^9$ & 6000\\
			5 & $-8.4686(7/8) \times 10^7$ & $2.616 \times 10^8$&$6.8 \times 10^{9}$ & 8000\\
			6 & $-2.145944 \times 10^8$ & $8.930 \times 10^8$ & $2.55 \times 10^{10}$ & 8000\\
			7 & $-4.69624 \times 10^8$ & $2.459 \times 10^9$ & $7.7(4\pm 2)\times 10^{10}$ & 8000\\
			8 & $-9.239(10\pm 1) \times 10^8$ & $5.831 \times 10^9$ & $2.01\times 10^{11}$ & 8000\\
			9 & $-1.67634 \times 10^9$ & $1.2384 \times 10^{10}$ & $4.65\times 10^{11}$ & 8000\\
			10 & $-2.85418 \times 10^9$ & $2.416 \times 10^{10}$ & $9.8(3\pm 2)\times 10^{11}$ & 8000\\
			15 & $-2.20129\times 10^{10}$ & $3.0492 \times 10^{11}$ & $1.731 \times 10^{13}$ & 8000\\
			20 & $-9.34553 \times 10^{10} $ & $1.7983 \times 10^{12}$ & $1.313 \times 10^{14}$ & 8000\\
			25 & $-2.8645\times 10^{11}$ & $7.056 \times 10^{12}$ & $6.31\times 10^{14}$ & 8000\\\hline 
		\end{tabular}
	\caption{The numerical values for the coefficients $C_i$ with different $k$. The range of our discretization points is $[0,L]$. }
		\label{DataDeltakLargeLambda}
	\end{table}

\textit{Note added:}
	While we were finalizing version 1 of this manuscript, \cite{Beccaria:2022ypy} appeared in which the large $\lambda$ expansion of $\Delta_k(\lambda)$ is calculated analytically. The result in \cite{Beccaria:2022ypy} agrees with our conjecture~\eqref{NikolayConjecture}. Their expansion also agrees with the expressions in~\eqref{C5p2} and~\eqref{C3}. Additionally, they obtained the closed form expressions
\begin{equation}
\epsilon_{5/2}  = \frac{1}{3} - \frac{\zeta(3)}{ 16 \log^3 2 }
\qquad \text{and} \qquad
\epsilon_{3}  = \frac{5}{3} - \frac{\zeta(3)}{ 2 \log^3 2 }\, ,
\end{equation}
which agree with our numerical estimates discussed above. It turns out that our numerics was not accurate enough to determine the full structure of the coefficient $C_{7/2}$ in~\eqref{C7p2}. Namely, instead of the factor in the square brackets $[(1+\epsilon_{7/2})k^2 - (2-2\epsilon_{7/2})]$, the analytical calculation in \cite{Beccaria:2022ypy} yields
\begin{equation}
	\label{neweqC72}
	\begin{aligned} 
		& \left(-\frac{9 \zeta_5 }{128 \log^52}+\frac{\zeta_3 }{\log ^32}-\frac{32}{15}\right) k^2  +
		\left(-\frac{9 \zeta_5 }{128 \log^52}-\frac{3 \zeta_3 }{4 \log
			^32}+\frac{16}{5}\right) k \\
		& \qquad + \left( -\frac{27 \zeta_5 }{2048 \log^52}-\frac{\zeta_3 }{4 \log
			^32}-\frac{16}{15} \right) \\
		& \approx [ 1.02051 k^2 +0.0371916 k-2.05448]\,.
	\end{aligned}
\end{equation}
The rest of the expression for $C_{7/2}$ in~\eqref{C7p2} agrees with the analytic result in \cite{Beccaria:2022ypy}.

\section{Free energy ${\cal F}$ and untwisted correlators}
\label{sec:FreeEnergy}

Our goal in this section is to compute the free energy of the matrix model at hand. To this end we employ the Weinstein-Aronszajn determinant identity for an $m \times n$ matrix $A$ and an $n \times m$ matrix $B$ which reads
\begin{equation} 
	\det(\mathbf{1}_{m\times m} + AB) = \det(\mathbf{1}_{n\times n} + BA)\,.
	\label{TaoMagicalFormula}
\end{equation} 
We can identify the matrix $\mathsf{X}$ defined in \eqref{Xx} as the product of two $A_k(t)$ which can be regarded as a generalized matrix with one discrete index $k\in \mathbb{N_+}$ and one continuous index $t\in (0,\infty)$. After the change of variables $t \rightarrow \frac{2\pi t}{\sqrt{\lambda}}$ we can rewrite \eqref{Xx} as
\begin{equation}
	\mathsf{X}_{kl} = -\int dt A_k(t)A_l(t),\quad A_k(t)\equiv (-1)^k \sqrt{(2k+1)\ \frac{16 \pi}{\sqrt{\lambda}}\ W\left( \frac{2\pi t}{\sqrt{\lambda}} \right)} J_{2k+1}(t)\,.
\end{equation}
Using a continuous version of formula (\ref{TaoMagicalFormula}), one can transform the free energy difference ${\cal F}$ defined in \eqref{FreeEnergyDifference} as follows:
\begin{equation}
	\begin{aligned}
		{\cal F} &= \frac{1}{2} \log \det (\mathbf{1}-\mathsf{X}) =  \frac{1}{2} \log \det \left( \mathbf{1} + \int dt A_k(t)A_l(t) \right)\\
		&\qquad  =  \frac{1}{2} \log \det \left( \mathbf{1} +\sum_k A_k(t) A_k(s) \right) \,.
	\end{aligned}
\end{equation}
We therefore conclude that ${\cal F}$ is the logarithm of a Fredholm determinant: 
\begin{equation}	\label{FredholmDeterminantF}
	{\cal F}= \frac{1}{2} \log \det( \mathbf{1} + K)\,,
\end{equation}
where $\mathbf{1}+ K$ is the functional operator that we encountered in \eqref{eq:eq7}. This will be our starting point for the calculation of the free energy. 

As a consistency check on the manipulations performed above, we can compare our results derived from \eqref{FredholmDeterminantF} against the small  $\lambda$ expansion derived in \cite{Tseytlin} by other methods.  In Section~\ref{sec:smallDeltaAN}, we showed that when performing an expansion for small $\lambda$ the kernel is degenerate. Therefore, we can calculate the Fredholm determinant in \eqref{FredholmDeterminantF} using Equation \eqref{eq:deg10}. To this end we employ the results and notation of Section~\ref{sec:smallDeltaAN}. For the free energy up to order $\mu^{10}$ we find
\begin{equation}
{\cal F} = \dfrac{1}{2} \log\det(1 + A)+O\left(\mu\right)^{10}\,,
\end{equation}
where the $3 \times 3$ matrix $A$ is given in equation \eqref{small:A}. This in turn leads to
\begin{equation}
{\cal F } = \frac{5 }{8}\zeta (5)\mu ^6 -\frac{105 }{32}\zeta (7)\mu ^8 +O\left(\mu\right)^{10}\,.
\end{equation}
This result agrees with the first two terms in Equation (3.4) in \cite{Tseytlin}. It is straightforward but tedious to extend our method to higher orders in $\mu$. In particular we have successfully compared our results to the terms up to order $\mu^{20}$ in Equation (3.4) of \cite{Tseytlin}.\footnote{All the terms agree, apart from one. In \cite{Tseytlin} there is a term $\frac{3.6355}{8} \zeta(13) \hat{\lambda}^7$, but we find instead $\frac{212355}{8} \zeta(13) \hat{\lambda}^7$. Given that $\frac{3.6355}{8} \zeta(13) \hat{\lambda}^7$ is the only term with irrational coefficient multiplying a $\zeta$ function in Equation (3.4) in \cite{Tseytlin}, this is probably a typo.}

\subsection{Numerical method for the calculation of ${\cal F}$}\label{sec:Fnum}

\begin{figure}[H]
	\centering
	\includegraphics[scale = 1]{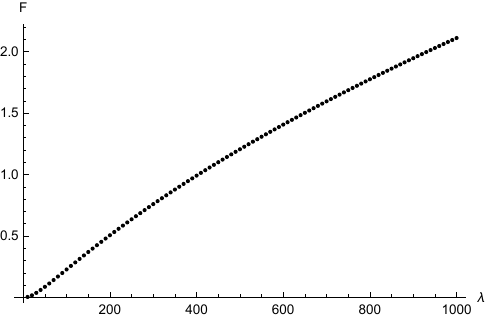}
	\caption{\rm $\cF$ as a function of $\lambda$ computed with the Bornemann method.}
	\label{FLambda}
\end{figure}

\begin{figure}[H]
	\centering
	\includegraphics[scale = .9]{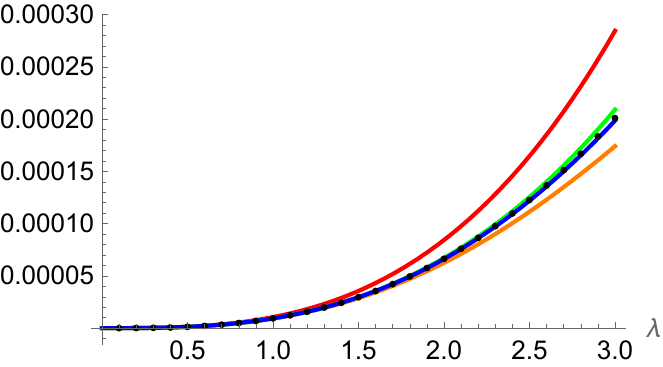}
	\caption{Black dots: numerical calculation of ${\cal F}$ with the Bornemann method. Red line: $ F_3 \lambda^3$; orange line: $ F_3 \lambda^3 + F_4 \lambda^4$; green line: $F_3 \lambda^3 + F_4 \lambda^4 + F_5 \lambda^5$; blue line: $F_3 \lambda^3 + F_4 \lambda^4 + F_5 \lambda^5 + F_6 \lambda^6 $. Clearly the numerical and analytic results agree very well.}\label{Figure0630b}
\end{figure}

Unfortunately we are not able to analytically calculate the Fredholm determinant in \eqref{FredholmDeterminantF} for all values of $\lambda$. To proceed we calculate ${\cal F}$ numerically with the Bornemann method~\cite{Bornemann}. This numerical method is based on the Nystr\"om method for solving integral equations. Namely, if $w_a \ge 0$ are weights and $t_a$ are discretization points of a quadrature method, with $a = 1, \ldots, m$, then
\begin{equation}\label{Bornemann}
\det( \mathbf{1} + K) \approx \det_{a,b=1}^m \Big( \delta_{ab} + \sqrt{w_a} K(t_a,t_b) \sqrt{w_b} \Big)\,.
\end{equation}
This algorithm is simple but very efficient for smooth kernels. There are many discretization schemes one can use in \eqref{Bornemann}. As in Section~\ref{sec:Deltanum} we have chosen Fej\'er type 1. 
As an illustration of the results obtained via this method, in Figure~\ref{FLambda} we present the numerical results for the free energy with some appropriate settings for the quadrature parameters $L$ and $m$.
A table with numerical values of  ${\cal F}$ can be found in Appendix~\ref{app:tables}. As a consistency check on the numerical implementation of the Fredholm determinant algorithm we can compare the numerical results against the analytic small $\lambda$ expansion. We use the small $\lambda$ expansion of Equation (3.4) in \cite{Tseytlin} which we have independently reproduced:
\begin{equation}\label{eq:Fsmall}
	{\cal F} = F_3 \lambda^3 + F_4 \lambda^4 + F_5 \lambda^5 + F_6 \lambda^6 + \cdots
\end{equation}
with  $F_3 = \dfrac{5}{512 \pi ^6}\zeta (5)$, $F_4 = -\dfrac{105 }{8192 \pi ^8}\zeta (7)$ etc. The results are illustrated in figure \ref{Figure0630b}.
In Table~\ref{tab:smallFlambda} we provide more evidence for the validity of our numerical method. As discussed in \cite{Tseytlin}, the radius of convergence of $\lambda$ is $\lambda_c = \pi^2$, which is the same as for $\cN = 4$ SYM.
\begin{table}[H]
	\centering
	$$
	\begin{array}{|l|lllll|}
		\hline
		n \backslash \lambda & 1 & 2 & 3 & 4 & 20 \\
		\hline
		1 & 0.0000105329 & 0.0000842635 & 0.000284389 & 0.000674108 & 0.0842635 \\
		2 & 0.00000917083 & 0.0000624697 & 0.000174058 & 0.000325408 & -0.133674 \\
		3 & 0.00000931482 & 0.0000670777 & 0.00020905 & 0.000472864 & 0.327125 \\
		4 & 0.00000930041 & 0.0000661549 & 0.000198539 & 0.000413804 & -0.595688 \\
		5 & 0.00000930182 & 0.0000663362 & 0.000201637 & 0.00043701 & 1.2173 \\
		10 & 0.0000093017 & 0.0000663066 & 0.000200935 & 0.000430498 & 249.84 \\
		15 & 0.0000093017 & 0.0000663066 & 0.000200935 & 0.000430495 & 960.633 \\
		\hline
		\text{numerics} & 0.0000093017 & 0.0000663066 & 0.000200935 & 0.000430494 &
		0.0175565 \\
		\hline
	\end{array}
	$$
	\caption{Comparison of small $\lambda$ expansion against numerical calculation of the free energy for different values of $\lambda$. The last row gives the value calculated with the Bornemann method. The first row with $n=1$ gives the result using 1 term in the expansion \eqref{eq:Fsmall}, the second row with $n=2$ gives the result using two terms in the expansion \eqref{eq:Fsmall} and so on. One can see that for small $\lambda < \lambda_c$ ($\lambda = 1,2,3,4$ in the table) the series \eqref{eq:Fsmall} converges to the numerical result as $n$ increases. For large $\lambda > \lambda_c $ ($\lambda = 20$ in the table), the series \eqref{eq:Fsmall} does not seem to converge, but the numerical method still yields sensible results.}\label{tab:smallFlambda}
\end{table}

\subsection{Large $\lambda$ expansion of ${\cal F}$: a conjecture}

Using the numerical method detailed above we calculated ${\cal F}$ for $\lambda$ going up to $e^{18}$. Fitting the numerical data listed in table \ref{TblLambdaF}, we observed that the following asymptotic expansion with coefficients in closed form agrees very accurately with the data
\begin{equation}
	{\cal F} = \frac{1}{8} \sqrt{\lambda}- \frac{3}{8} \log \left(\frac{\lambda}{\lambda_0}\right) - \frac{3}{4} \frac{\log 2}{\sqrt{\lambda}} 
	- \frac{3}{2} \frac{\log^2 2}{\lambda} + O( \lambda^{-3/2})\,.
	\label{FF}
\end{equation}
The constant $\lambda_0 \approx 7.723901172$ can be determined with high precision but we were not able to identify it with a closed form irrational or transcendental number. It is tempting to speculate that the coefficient of the $\log\lambda$ term above is related to the difference in conformal anomalies between the $\mathbf{E}$ theory and $\mathcal{N}=4$ SYM, see \eqref{eq:acEN4}, but we were not able to make this statement quantitatively precise. 

\begin{table}[h]
	\centering
	\begin{tabular}{|c||c|c|c|c|}
		\hline
		$\lambda$ & $\cF$  \\\hline\hline 
		$e^8$ & $4.5816147173   (9\! \pm \! 6 )$\\
		$e^{8.5}$ & $ 6.3347294273322   (8\! \pm \! 8 ) $\\
		$e^{9}$ & $ 8.637894890881   (0\! \pm \! 9 ) $ \\
		$e^{9.5}$ & $ 11.64760271190   (0\! \pm \! 6 ) $ \\
		$e^{10}$ & $ 15.5647287113   (94\! \pm \! 21 ) $\\
		$e^{10.5}$ & $ 20.6471553368   (9\! \pm \! 4 )$\\
		$e^{11}$ & $ 27.225974672885   (68\! \pm \! 15 )$\\
		$e^{11.5}$ & $ 35.726290419555   (6\! \pm \! 8 )$\\
		$e^{12}$ & $ 46.69392596735   (85\! \pm \! 24 )$ \\
		$e^{12.5}$ & $ 60.8297166581   (16\! \pm \! 11 )$\\
		$e^{13}$ & $ 79.033540756198   (14\! \pm \! 11 )$\\
		$e^{13.5}$ & $ 102.460855469831   (90\! \pm \! 14 )$\\
		$e^{14}$ & $ 132.595289990373   (00\! \pm \! 26 )$\\
		$e^{14.5}$ & $ 171.34185630862   (66\! \pm \! 11 )$\\
		$e^{15}$ & $ 221.1466338982   (52\! \pm \! 31 )$\\
		$e^{15.5}$ & $ 285.1504476051873   (00\! \pm \! 31 )$\\
		$e^{16}$ & $ 367.3861937408442   (07\! \pm \! 28 )$\\\hline 
	\end{tabular}
	\caption{The numerical values of $\cF$ that we used for numerical fitting. The range of our sampling points is $(0,L)$, where in practice we take $L = 12000$. }
	\label{TblLambdaF}
\end{table}

\begin{figure}
	\centering
	\includegraphics[width=12cm]{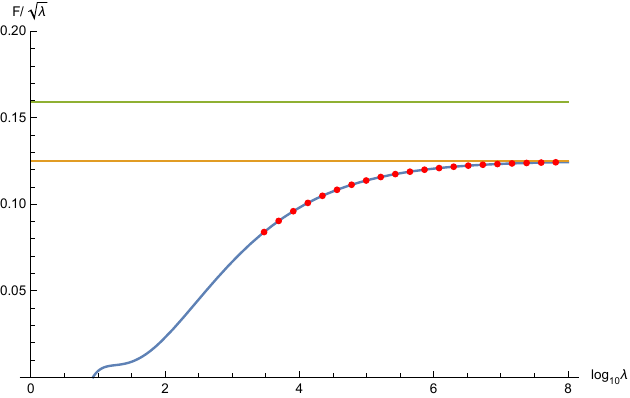}
	\caption{\rm The counterpart of Figure 4 in \cite{Tseytlin}. The green and dark yellow lines represent the constants $1/2\pi$ and $1/8$, respectively. The blue line is the analytic approximation in \eqref{FF}. The red dots are the numerical results obtained by evaluating the Fredholm determinant with the Bornemann method.}
	\label{LambdaExp}
\end{figure}

Interestingly the result in \eqref{FF} is in conflict with the leading order result presented in \cite{Tseytlin} where it was argued that $\cF \sim \frac{1}{2\pi}\sqrt{\lambda}$. As illustrated in Figure~\ref{LambdaExp} this is in clear disagreement with our numerical results. While we are not completely convinced why the result in  \cite{Tseytlin} disagrees with ours we suspect that it could be due to the assumption in \cite{Tseytlin} that the leading order behavior of $\cF$ for large $\lambda$ is controlled by the leading order behavior of $\mathsf{X}$. Namely, in \cite{Tseytlin} the authors use\footnote{The expression of matrix $\mathsf{S}$ is given in \eqref{s}.}
\begin{equation}\begin{aligned}
		\cF &= \frac{1}{2}\log \det \left(\mathbf{1} - \mathsf{X}\right) = \frac{1}{2}\log \det \left[ \mathbf{1} + \frac{\lambda}{2\pi^2} \mathsf{S} + O(\lambda^0) \right]\\
		&\quad  =  \frac{1}{2}\log \det \left( \mathbf{1} + \frac{\lambda}{2\pi^2}\mathsf{S} \right) + o(\sqrt{\lambda})\,.
	\end{aligned}
	\label{ConjectureRelation}
\end{equation}

Both analytical and numerical analysis supports the following behavior, see \cite{Tseytlin} 
\begin{equation}
	\frac{1}{2}\log \det \left( \mathbf{1} + \frac{\lambda}{2\pi^2}\mathsf{S} \right) \sim \frac{1}{2\pi} \sqrt{\lambda} 
	\label{LeadingOrderApproximation}
\end{equation}
Thus we suspect that there is something wrong in going from the first to the second line in \eqref{ConjectureRelation} which is the cause of the conflict with our numerical analysis. To illustrate this subtlety we have shown numerically that $\cF/\sqrt{\lambda}$ can also depend on the sub-leading order expansion of $\mathsf{X}$, for example, if we multiply the identity matrix in \eqref{LeadingOrderApproximation} with an arbitrary constant, then the coefficient of $\sqrt{\lambda}$ in the final result changes. As discussed in Appendix \ref{sec:term2} similar subtle issues are present in the analytic evaluation of $\Delta_k$.

\textit{Note added:} While we were finalizing version 1 of this manuscript, \cite{Beccaria:2022ypy} appeared in which the large $\lambda$ expansion of ${\cal F}$ is calculated analytically. The result in \cite{Beccaria:2022ypy} agrees with equation~\eqref{FF}. The authors of \cite{Beccaria:2022ypy} also calculated that
\begin{equation}\label{eq:lambda0v2}
	\lambda_0 = 2^{ 14/9} e^{ 8 \zeta'(-1)} \pi^2\, ,
\end{equation}
which agrees with the numerical value we found.

\subsection{Untwisted correlators and Wilson loops in the $\mathbf{E}$ theory}\label{FSection}

The free energy in \eqref{FF} can be used to calculate the large $\lambda$ behavior of the functions $\delta_k(\lambda)$ that control the two-point functions of the untwisted sector operators  in the $\mathbf{E}$ theory, see \eqref{DeltaAnddelta}. It was argued in \cite{Billo1} that $\delta_k(\lambda)$ obey the following relation 
\begin{equation}\label{eq:deltakrelF}
	\delta_k(\lambda) = -2k\left[ (2k^2-1)\lambda\partial_\lambda \cF + (\lambda\partial_\lambda)^2\cF \right],
\end{equation}
	where $\cF$ is the free energy defined in \eqref{FreeEnergyDifference}. Using \eqref{eq:deltakrelF} and the conjectured analytic form of the large $\lambda$ behavior of $\mathcal{F}$ in \eqref{FF} we find that the large $\lambda$ behavior of $\delta_k(\lambda)$ is 
\begin{multline}
\delta_k(\lambda) =  -\frac{k(4k^2-1) }{16}\sqrt{\lambda} + \frac{3}{8}k(4k^2-2)\\- \frac{3}{8}\frac{k(4k^2-3)\log 2}{\sqrt{\lambda}} -\frac{3}{2}  \frac{k(4k^2-4) \log^22}{\lambda}+O(\lambda^{-3/2})\,.
\end{multline}
This is a new result that provides three more orders in the large $\lambda$ expansion of $\delta_k(\lambda)$ as compared to the previous literature \cite{Billo1,Tseytlin}. Note that we again find a discrepancy between our result for the coefficient of the $\sqrt{\lambda}$ term in $\delta_k(\lambda)$ above and the results of \cite{Tseytlin}. This is due to the discrepancy in $\mathcal{F}$ discussed around \eqref{ConjectureRelation}.

As a final application of our result for $\mathcal{F}$ we can calculate the first four terms in the strong coupling expansion of the function $q(\lambda)$ that controls the vacuum expectation value of a 1/2-BPS circular expectation loop in the $\mathbf{E}$ theory. As derived in \cite{Tseytlin} and discussed around equations \eqref{eq:Deltaqdef} and \eqref{eq:Deltaqres} this function can be calculated by taking a derivative of $\mathcal{F}$, $\Delta q (\lambda) = - \frac{\lambda^2}{4} \partial_{\lambda} \mathcal{F}(\lambda)$. Using our result for $\mathcal{F}$ in \eqref{FF} we find 
\begin{equation}
\Delta q (\lambda) = - \frac{1}{64} \lambda^{3/2}+\frac{3}{32} \lambda - \frac{3\log2}{32} \lambda^{1/2} - \frac{3\log^22}{8} +O(\lambda^{-1/2})\,.
\end{equation}
%

\section{Three point functions}
\label{sec:Threeptfc}

In this section, we study the strong coupling expansion of three-point functions of single trace operators in the  {\bf E} theory. As discussed in Section \ref{sec:N2summary}, $G_{m,n}$, which are defined in equation~\eqref{FTCorrelators}, can be calculated with an expectation value in the matrix model. In the planar limit this is:
\begin{equation}
		G_{m,n} =  \langle O_m(a) O_n(a) O_{m+n}(a) \rangle\\\,.
\end{equation} 
Its value in  $\cN = 4$ YM is given by, see for example \cite{Billo:2022xas},
\begin{equation}\label{GmnN=4}
	\quad G_{m,n}^{(0)} = \frac{mn(m+n)}{2}\left( \frac{N}{2}\right)^{m+n-1}\,.
\end{equation}
We will calculate $\Delta_{m,n}(\lambda)$ which is defined by
\begin{equation}
	G_{m,n} = G_{m,n}^{(0)} \left( 1 + \Delta_{m,n}(\lambda) + O(1/N)^2 \right).
\end{equation}
	There are three possibilities for the parities of $m$ and $n$. If $m$ and $n$ are both even, the three point functions belong to the un-twisted sector, and therefore in the planar limit $\Delta_{2k, 2l} = 0$. 
	For the cases where $m$ and $n$ are both odd, or one is odd and the other is even, $\Delta_{m,n}$ is generally non-zero. Their leading order behaviour at large $\lambda$ was calculated previously in \cite{Billo:2022xas}. In the next few sections we will calculate the next three terms in its expansion. The reader who is not interested in the calculation can find the result in equations~\eqref{Xuao3pt1} and~\eqref{Xuao3pt2} below. We want to emphasize that the calculation below is fully analytic, but we use the 
	conjectured form~\eqref{Dkl} for  $\mathsf{D}_{kl}$ as input.

Before calculating $\Delta_{m,n}$ we need to calculate several auxiliary objects called $T$, $C$ and $M$. The $T$'s are defined by 
the expectation values
	\begin{equation}
	T_{m}  = \langle \Omega_{m}(a) \rangle ,\quad T_{m,n}  = \langle \Omega_{m}(a) \Omega_{n}(a) \rangle ,\quad\text{etc}\,.
\end{equation} 
The mixing coefficients $C_{m,n}$ were defined previously in equation~\eqref{mixingCmn}. The matrix $M$ is given by inverting a matrix: $M = (\mathbf{1} + C)^{-1}$. We will first present the expressions for $T$, $C$, $M$ in the untwisted sector and twisted sectors. These results can be found in \cite{Billo:2022xas} but we include them here because we need them in Section~\ref{3ptFunction} where we put everything together and calculate the strong coupling expansion of $\Delta_{m,n}(\lambda)$.
	The two-point functions of operators with even dimensions belong to the un-twisted sector, therefore\footnote{All equations presented in this section are valid at the leading order in the large $N$ expansion.}:
	\begin{equation}\label{TEven}
		T_{2k,2l} = T_{2k,2l}^{(0)} = \frac{N^{k+l+2}}{2^{k+l}} \frac{(2k)!(2l)!}{k!(k+1)!l!(l+1)!} \,.
	\end{equation}
	In $\cN = 4$ YM, the mixing coefficients are
	\begin{equation}\label{CnmN=4YM}
		C_{n,m}^{(0)} = \lim\limits_{\lambda\rightarrow 0} C_{n,m} = \left( \frac{N}{2} \right)^{\frac{n-m}{2}}  \left( \begin{array}{c} n \\ \frac{n-m}{2}  \\ \end{array}\right),\quad \text{if } n > m,
	\end{equation}
and $C_{n,m}^{(0)} = 0$ if $n\le m$, thus for the un-twisted sector, we have
	\begin{equation} \label{CMatrixEven}
		C_{2k, 2l} =  C_{2k,2l}^{(0)} = \left(\frac{N}{2}\right)^{k-l}   \left( \begin{array}{c} 2k \\ k - l \\ \end{array}\right),\quad\text{if }k>l.
	\end{equation}
	We introduce the vev-less version of $\Omega_n(a)$:
	\begin{equation}
		\widehat{\Omega}_n(a) \equiv \Omega_n(a) - \langle \Omega_n(a) \rangle = \Omega_n(a) - T_n,
	\end{equation}
	and we denote the transformation coefficients from $\widehat{\Omega}_n(a)$ to $O_n(a)$ as $M_{n,m}$: 
	\begin{equation} \label{ExpandOn}
		O_n(a) = \sum_{m \le n} M_{n, m } \widehat{\Omega}_m (a).
	\end{equation}
	Using~\eqref{O2Omega} on finds: 
	\begin{equation}
		M_{n,m} = \left( \frac{1}{\mathbf{1} + C} \right)^{-1}_{n,m}\,.
	\end{equation}

The matrices $M$ and $C$ are lower-triangular, so we obtain from~\eqref{CnmN=4YM}
	\begin{equation}\label{MnmN=4YM}
		M_{n,m}^{(0)} = \left( - \frac{N}{2} \right)^{\frac{n-m}{2}} \frac{n}{m} \left( \begin{array}{c} \frac{n+m-2}{2} \\ \frac{n-m}{2} \\\end{array}\right),\quad\text{if } n\ge m\, ,
	\end{equation}
	and $M_{n,m}^{(0)} = 0$ if $n < m$. Therefore in the un-twisted sector we have 
	\begin{equation}
	M_{2k, 2l} = M_{2k, 2l}^{(0)} = \left( - \frac{N}{2} \right)^{k-l} \frac{k}{l} \left( \begin{array}{c} k+l-1\\ k-l\\\end{array}\right) ,\quad \text{if } k\ge l.
	\end{equation}
		When $\lambda = 0$, one has \cite{RodriguezGomez:2016wl}:
		\begin{equation}\label{Chebyshev}
				O_n^{(0)}(a) = {\rm tr}\, p_n(a),\quad p_n(a)\equiv 2 \left(\frac{N}{2}\right)^{\frac{n}{2}} T_n \left( \frac{a}{\sqrt{2N}} \right)+ \delta_{n,2} \mathbf{1} = a^n + \cdots\,.
		\end{equation}
		where $T_n(x)$ is the Chebyshev polynomial of the first kind. 
	
		The two-point functions $T_{2k+1, 2l+1}$ can be evaluated using $\mathsf{D}_{k,l}$. From  \eqref{MnmN=4YM} and \eqref{Chebyshev}, one finds
	\begin{equation}
		\Omega_{2k+1}(a) = \left( \frac{N}{2} \right)^{k+ \frac{1}{2}} \sum_{i=0}^{k-1} c_{k,i} \omega_{k-i}(a),\quad\text{with}\quad c_{k,i}=\left( \begin{array}{c} 2k+1 \\ i \\\end{array}\right)\sqrt{2k-2i+1},
	\end{equation}
	and therefore for $k,l \ge 1$
	\begin{equation}\label{TOdd}
		T_{2k+1, 2l+1}  =  \left( \frac{N}{2} \right)^{k+l + 1} \sum_{i=0}^{k-1}\sum_{j=0}^{l-1} c_{k,i}c_{l,j} \mathsf{D}_{k-i,l-j}\,.
	\end{equation}

	Using our conjectured expression~\eqref{Dkl} for $\mathsf{D}_{kl}$ it is then straightforward to show that
		\begin{equation} \label{TOddExp}
			 \begin{aligned}
				T_{2k+1,2l+1} =&  \left(\frac{N}{2} \right)^{k+l+1} 4\pi^2 \frac{(2k+1)!(2l+1)!}{k!(k-1)!l!(l-1)!}\left[\frac{1}{k+l} \dfrac{1}{\lambda}  - \frac{8\log 2}{\lambda^{3/2}} \right.\\
				& 	\left.  +  \frac{16\log^22}{\lambda^2} (2k+2l-1) + \frac{4\pi^2}{3\lambda^2} \frac{k(k-1)+l(l-1)}{k+l-1} \right.\\
				& \left. - \frac{32}{3} \frac{\log2}{\lambda^{5/2}}  \left(\pi ^2 (k(k-1)+l(l-1)) + 4 \log^22 (k+l-1) (2 k+2 l-1)\right) \right.\\
				& \left. + \frac{\zeta(3)}{\lambda^{5/2}} (-32 k^2 -32 l^2 + 24 k + 24 l + 32 k l) + O(\lambda)^{-3}\right]\,.
		\end{aligned}
	\end{equation}
		The mixing coefficients $C_{2k+1, 2l+1}$ for $k > l \ge 1$ can be obtained from $T_{2k+1, 2l+1}$ using the following formula \cite{Billo:2022xas}: 
	\begin{equation}\label{COddFromTOdd}
		C_{2k+1, 2l+1} = \left| \begin{array}{cccc}  T_{3,3} & T_{3,5} & \cdots & T_{3, 2k+1} \\ T_{5,3} & T_{5,5} & \cdots & T_{5,2k+1}\\ \vdots & \vdots & \ddots & \vdots \\ T_{2l+1,3} & T_{2l+1,5} & \cdots & T_{2l+1, 2k+1} \end{array} \right| \Big/		
		\left| \begin{array}{cccc}  T_{3,3} & T_{3,5} & \cdots & T_{3, 2l+1} \\ T_{5,3} & T_{5,5} & \cdots & T_{5,2l+1}\\ \vdots & \vdots & \ddots & \vdots \\ T_{2l+1,3} & T_{2l+1,5} & \cdots & T_{2l+1, 2l+1} \end{array} \right| \ .
	\end{equation}
	Here the denominator is the determinant of an $l\times l$ matrix $T_{2i+1, 2j+1}$ with $1\le i,j \le l$, and the matrix in the numerator is obtained by performing in the last column the replacement $l \rightarrow k$. 
	Substituting~\eqref{TOddExp} in~\eqref{COddFromTOdd} one obtains
	\begin{equation} \label{CMatrixOdd}
	\begin{aligned}
			C_{2k+1, 2l+1} =&  \left(\frac{N}{2}\right)^{k-l} \frac{2k+1}{2l+1}  \left( \begin{array}{c} 2k \\ k - l \\ \end{array}\right) \left[ 1 - \frac{8\log 2}{\lambda^{1/2}} (k-l) + \right.\\
			& \left. + \frac{16 \log^22 }{\lambda} (k-l) (2k-2l-1) + \frac{4\pi^2}{3\lambda}(k+l)(k-l)\right.\\
			& \left. -\frac{32\log 2}{3\lambda^{3/2}} (k-l-1) (k-l) \left(4 \log ^22 (2 k-2 l-1)+\pi ^2  (k+l)\right)\right.\\
			& \left. + \frac{\zeta (3)}{\lambda^{3/2}} (k-l) \left(-32 k^2 +16 l^2 -32 k l+24 k+24 l-1 \right) + O(\lambda)^{-2}\right]\,.
	\end{aligned} 
	\end{equation}
		Inverting $\mathbf{1} + C$ with odd indices we then find that for $ k \ge l$
	\begin{equation}\label{MOdd}
		\begin{aligned}
				M_{2k+1, 2l+1} =& \left( -\frac{N}{2}\right)^{k-l} \frac{2k+1}{2l+1} \frac{k}{l}  \left( \begin{array}{c} k+l -1\\ k - l\\\end{array}\right) \left[ 1 - \frac{8\log2}{\lambda^{1/2}} (k-l) \right.\\
				& \left. + \frac{16\log^22}{\lambda}(k-l)(2k-2l-1) + \frac{4\pi^2}{3\lambda} (k+l)(k-l) \right.\\
				& \left. -\frac{32\log 2}{3\lambda^{3/2}} (k-l-1) (k-l) \left(4 \log ^22 (2 k-2 l-1)+\pi ^2  (k+l)\right)\right.\\
				& \left. + \frac{\zeta (3)}{\lambda^{3/2}}  (k-l) \left(16 k^2 -32 l^2 -32 k l-24 k-24 l-1 \right) + O(\lambda)^{-2} 
				\right]\,.
		\end{aligned}
		\end{equation}
		
The last quantity we will need is the three-point functions of $\Omega_n(a)$. This can be found in \cite{Billo:2022xas} and reads:
		\begin{equation} \label{Tmnp}
			T_{2m, 2k+1, 2l+1} = T_{2m} \left[ 1 + \frac{m(m+1)}{N^2} \left( k+l+1 + \lambda \partial_\lambda \right) \right] T_{2k+1, 2l+1} + O(1/N)^4.
		\end{equation}

	\subsection{Three point function at large $\lambda$ } \label{3ptFunction}
	Using \eqref{ExpandOn}, we have
		\begin{equation}\label{3ptFunctionRaw}
			\begin{aligned}
				\langle O_{2m}(a) &  O_{2l+1}(a) O_{2p+1}(a) \rangle = \sum_{(n,r,s)=1}^{(m,l,p)} M_{2m, 2n}M_{2l+1,2r+1}M_{2p+1, 2s+1} \langle \widehat{\Omega}_{2n}\widehat{\Omega}_{2r+1} \widehat{\Omega}_{2s+1} \rangle \\
				&=  \sum_{(n,r,s)=1}^{(m,l,p)} M_{2m, 2n}M_{2l+1,2r+1}M_{2p+1, 2s+1}(T_{2n, 2r+1, 2s+1} - T_{2n}T_{2r+1, 2s+1})\,.\\
			\end{aligned}
		\end{equation}
			On the second line above, we substituted the definition of $\widehat{\Omega}_n$ and simplified.  Using \eqref{Tmnp}, the term in the bracket is given, in the large $N$ limit, by 
		\begin{equation}
			T_{2n, 2r+1, 2s+1} - T_{2n}T_{2r+1, 2s+1} = \frac{n(n+1)}{N^2}T_{2n} \left[ (r+s+1) T_{2r+1, 2s+1} + \lambda \partial_\lambda T_{2r+1, 2s+1}  \right].
		\end{equation}
		Using~\eqref{3ptFunctionRaw}, we find that the summation of $n$ can be performed explicitly. Namely, using the expressions for  $T_{2n}$ and $M_{2m, 2n}$ in \eqref{TEven} and \eqref{MnmN=4YM}, one has
		\begin{equation}
				\sum_{n=1}^m M_{2m, 2n} T_{2n} \frac{n(n+1)}{N^2} = m \left(\frac{N}{2} \right)^{m-1}\,.
		\end{equation}
		The sum over $r, s$ can also be done directly using \eqref{TOddExp} and \eqref{MOdd}. Putting everything together, we find
		\begin{equation}\label{3ptFunctionGeneral}
		\begin{aligned}
					\langle O_{2m} O_{2l+1} O_{2p+1} \rangle =&  \left(\frac{N}{2}\right)^{p+l+m}  \frac{16\pi^2m}{\lambda} l(2l+1)p(2p+1) \left[ 1 - \frac{4 \log 2}{\lambda^{1/2}} (2p+2l-1 )\right. \\
					&\left.\  + \frac{16\log^22}{\lambda} (2p+2l-1)(p+l-1)\right.\\
					&\left.\ - \frac{64\log^32}{3\lambda^{3/2}}(2p+2l-1)(p+l-1)(2p+2l-3) \right.\\
					&\left.\  + \frac{\zeta (3) }{2\lambda^{3/2} } (2p+2l-1) \left( 16 p^2 + 16 l^2-16 pl-4 p-4 l-3 \right)+ O(\lambda)^{-2} \right]\,.
		\end{aligned} 
		\end{equation}
		If we take $m = k$ and $p = k+l$, use \eqref{Def3ptFunctions} and the value of $G_{m,n}^{(0)}$ in \eqref{GmnN=4}, 
		we finally find 
	\begin{equation}
	\label{Xuao3pt1}		
	\begin{aligned} 
		1 + \Delta_{2k,2l+1} =&  \frac{16\pi^2}{\lambda}  l(k+l) \left[ 1 -  \frac{4\log 2 }{\lambda^{1/2}} (2k + 4l -1)  + \frac{16\log^22}{\lambda} (2k+4l-1)(k+2l-1)\right. \\
		&\left. - \frac{64\log^32}{3\lambda^{3/2}}(2k+4l-1)(k+2l-1)(2k+4l-3) \right.\\
		&\left. + \frac{\zeta (3) }{2\lambda^{3/2} } (2k+4l-1) \left( 16 k^2 +16 l^2 +16 k l-4 k-8 l-3 \right)+ O(\lambda)^{-2} \right]\\
	\end{aligned}
\end{equation}
	If we take $m = k + l + 1$ and $p = k$ instead, we find
%
%
%
	\begin{equation}
	\label{Xuao3pt2}		
	\begin{aligned} 
	1 + \Delta_{2k+1,2l+1} =&  \frac{16\pi^2}{\lambda} kl \left[ 1  -  \frac{4\log 2 }{\lambda^{1/2}} (2k+2l-1 )  + \frac{16\log^22}{\lambda} (2k+2l-1)(k+l-1 ) \right. \\
	&\left. - \frac{64\log^32}{3\lambda^{3/2}}(2k+2l-1)(k+l-1)(2k+2l-3) \right.\\
	&\left. + \frac{\zeta (3) }{2\lambda^{3/2} } (2k+2l-1) \left( 16 k^2 + 16 l^2-16 kl-4 k-4 l-3 \right)+ O(\lambda)^{-2} \right]\\
\end{aligned}
\end{equation}
The leading term in the large $\lambda$ expansion in equations~\eqref{Xuao3pt1} and~\eqref{Xuao3pt2} agrees with the result in \cite{Billo:2022xas}, the next three subleading terms are our novel results based on the conjecture for $\mathsf{D}_{kl}$ in \eqref{Dkl}.

In \cite{Beccaria:2022ypy}, which appeared while the first version of this manuscript was being finished, it was pointed out that the expression for $\Delta_k(\lambda)$ simplifies when using the variable $\lambda'$, where	
\begin{equation}	
	( \lambda')^{1/2} = \lambda^{1/2} -4\log 2\,.
\end{equation}
We observe that something similar happens for $\Delta_{k,l}(\lambda)$, namely we have
\begin{multline}
		\label{Xuao3pt3}		
		1 + \Delta_{2k,2l+1}(\lambda) = 
		 \frac{16\pi^2}{\lambda}  l(k+l)\left( \frac{\lambda'}{\lambda} \right)^{(2k+4l-1)/2}\\
		 \left[ 1  + 
		\frac{\zeta (3) }{2(\lambda')^{3/2} } (2k+4l-1) \left( 16 k^2 +16 l^2 +16 k l-4 k-8 l-3 \right) 
		+ O(\lambda')^{-2} \right]\,,
\end{multline}
and
\begin{multline}
		\label{Xuao3pt4}				
		1 + \Delta_{2k+1,2l+1}(\lambda) =  \frac{16\pi^2}{\lambda} kl \left( \frac{\lambda'}{\lambda} \right)^{(2k+2l-1)/2}\\
		\left[ 1  + \frac{\zeta (3) }{2(\lambda')^{3/2} } (2k+2l-1) \left( 16 k^2 + 16 l^2-16 kl-4 k-4 l-3 \right)
		+ O(\lambda')^{-2} \right]\,.
\end{multline}

That this simplification also occurs for $\Delta_{k,l}(\lambda)$ seems to provide additional support for the conjectured form of $\mathsf{D}_{kl}$ in \eqref{Dkl}. Also, based on the form for $\Delta_k(\lambda)$ in \cite{Beccaria:2022ypy}, it is tempting to speculate that the expressions ~\eqref{Xuao3pt3} and~\eqref{Xuao3pt4} are correct up to $O(\lambda')^{-5/2}$ instead of only up to $O(\lambda')^{-2}$ displayed above, and that the coefficient of the next non-zero term is proportional to $\zeta(5)$.

\section{Discussion}
\label{sec:Discussion}

In this work, our main focus was on developing an efficient numerical algorithm that allows for the calculations of correlation functions in the $\mathbf{E}$ theory at strong coupling. To this end, we exploited recent supersymmetric localization results that reduce the calculations of these observables to matrix models that can be analyzed with a variety of techniques. Based on our numerical studies we were able to extract analytic expression for a few of the leading terms in the strong coupling expansion of these observables. Clearly, it is important to develop analytic techniques to derive these results more rigorously. To this end, we note that the kernel of the integral operator appearing in this paper is {\it integrable}~\cite{Deift, Korepin}. This type of kernel appears often in the mathematical physics literature. In that context, one often translates the calculation to a Riemann-Hilbert problem and the expansion for large parameters is carried out with a non-linear steepest descent method \cite{DeiftZhou}. Perhaps these methods can also be applied to our setup to calculate the large $\lambda$ expansion of the correlators analytically. 

The numerical techniques we developed can also be efficiently applied to calculate correlators at finite values of the coupling. This provides valuable lessons about the properties of large $N$ gauge theories at finite coupling. This can be used for instance to understand the convergence properties of the weak and strong coupling expansion and to gain insights into non-perturbative effects. We believe that very similar methods can be applied also to other examples of 4d $\mathcal{N}=2$ SCFTs that can be analyzed with supersymmetric localization. The Lagrangian theories discussed in Section~\ref{sec:N2summary} are natural candidates for such an analysis. Work along these lines has recently appeared in \cite{Billo:2022gmq, Billo:2022fnb,Galvagno:2020cgq,Galvagno:2021bbj} for a class of quiver gauge theories and we hope that our techniques will find an application and can be adapted to the study of these models.

The strong coupling results obtained here have connections to string theory through the AdS/CFT correspondence. It will be most interesting to understand how to calculate any of the correlators we analyzed by using world sheet methods. This problem appears to be highly non-trivial since it requires performing string world sheet calculations in the AdS$_5\times S^5/\mathbb{Z}_2$ orientifold at high order in the $\alpha'$ perturbative expansion. We hope that our results will serve as an impetus to tackle this challenging world sheet analysis.

\subsection*{Acknowledgments}
This work is supported in part by an Odysseus grant G0F9516N from the FWO and by the KU Leuven C1 grant ZKD1118 C16/16/005. PJDS would like to thank M.~Bill\`o for interesting discussions during the initial stages of this work. We are also grateful to the authors of \cite{Beccaria:2022ypy} for useful correspondence after the first version of this work appeared.

\appendix

\section{Wrong analytical calculation of $\Delta_k(\lambda)$ for large $\lambda$}
\label{sec:expansionDeltak}

In this section we try to calculate analytically the first two terms in the large $\lambda$ expansion of $\Delta_k(\lambda)$. We warn the reader however that although the calculation appears to be quite straightforward, the result turns out to be wrong. 
	
\subsection*{Evaluation of the $\lambda^{-1}$ term in $\Delta_k(\lambda)$}\label{sec:term1}

The calculation is based on the $LDU$ decomposition of matrices and uses the following fact. If $A = L \mathbb{D} U$ where $L$ is a lower triangular matrix, $ \mathbb{D}$ is a diagonal matrix and  $U$ is an upper triangular matrix, then	\begin{equation}
		\label{eq:LDUblock}
		A_{(k)} = L_{(k)}  \mathbb{D}_{(k)} U_{(k)}\,,
	\end{equation}
 where $A_{(k)}$ is the upper left $k\times k$ block in the matrix $A$ and we have used the same notation for the other matrices.
	Furthermore, if $L$ is lower uni-triangular, i.e. $L$ has ones on the diagonal, and $U$ is upper uni-triangular, then
	\begin{equation}
		\label{ratiodet}
		\frac{ \det{A_{(k)}}}{\det{A_{(k-1)}}} = \mathbb{D}_{kk}\,,
	\end{equation}
	with the convention that $A_{(0)} = 1$. 

The $\lambda^{-1}$ term in the strong coupling expansion of $\Delta_k(\lambda)$ was calculated in \cite{Billo1} with a different method, we calculate it here again with formula \eqref{ratiodet}. By performing the Mellin transformation of \eqref{Xx}, one has \cite{Billo1}:
\begin{align}
	\label{xs}
	\mathsf{X} \underset{\lambda \to \infty}{\sim}
	- \frac{\lambda}{2 \pi^2}  \, \mathsf{S} + O(\lambda)^0
\end{align}
where $\mathsf{S}$ is a tri-diagonal infinite matrix whose elements are
\begin{align}
	\label{s}
	\mathsf{S}_{kl}
	=  \frac{1}{4} (-1)^{k+l} \sqrt{\frac{2l+1}{2k+1}} \,
	\Big(\frac{\delta_{k-1,l}}{k\,(2k-1)} + \frac{\delta_{k,l}}{k\,(k+1)} +
	\frac{\delta_{k+1,l}}{(k+1)\,(2k+3)}\Big) ~.
\end{align}
One can then show that
\begin{align}
	\label{Dll}
\Dx \equiv  \frac{1}{1-\mathsf{X}}  =  \frac{2\pi^2}{\lambda} \mathsf{S}^{-1} + O(\lambda)^{-3/2}\,.
\end{align} 
Now let us define a lower uni-triangular matrix $L$ and a diagonal matrix $\mathbb{D}$
\begin{equation}\label{eq:LDdef}
	L_{k l} =  
	\begin{cases}
		1 & \text{if} ~k =  l~\\[4mm]
		-\dfrac{ \sqrt{2 k +1}}{\sqrt{ 2 l+1}}& \text{if} ~k= l+1\\[4mm] 
		0& \text{else}
	\end{cases},\qquad 	\mathbb{D}_{k l} =  
\begin{cases}
\dfrac{\lambda}{4 \pi^2}\dfrac{1}{(2 k+1) 2 k }& \text{if} ~k = l~.\\[4mm] 
0& \text{else}
\end{cases}
\end{equation}
One can check explicitly that
\begin{equation}
L^T \mathbb{D} L = \frac{\lambda}{2 \pi^2}\ \mathsf{S}
\end{equation}
which in turn leads to
\begin{align}
	\label{eq1}
	L^{-1} \mathbb{D}^{-1} L^{-T} = \frac{2 \pi^2}{\lambda}\ \mathsf{S}^{-1} = \mathsf{D} + O(\lambda)^{-3/2}\,.
\end{align}
Since $L^{-1}$ is again a lower triangular matrix, one can use formula \eqref{ratiodet} to get:
\begin{align}
	1 + \Delta_k(\lambda) + O(\lambda)^{-3/2} =  \frac{\det \mathsf{D}_{(k)}}{\det \mathsf{D}_{(k-1)}} + O(\lambda)^{-3/2} = \mathbb{D}^{-1}_{kk}  + O(\lambda)^{-3/2}\,.
\end{align}
	Substituting the value of $\mathbb{D}_{kk}^{-1}$, we get:
\begin{align}\label{eq:approx1}
	1 + \Delta_k(\lambda) =  \dfrac{4 \pi^2}{\lambda}(2 k+1) 2 k +O(\lambda)^{-3/2}
\end{align}

This agrees with both the analytical evaluation in \cite{Billo1} and the numerical evaluation \eqref{NikolayConjecture} in the main text above.

\subsection*{Evaluation of the $\lambda^{-3/2}$ term in $\Delta_k(\lambda)$}\label{sec:term2}
To proceed to higher order in the $1/\lambda$ expansion we add one more term to the expansion of $\mathsf{X}$:
\begin{align}
	\label{x2}
	\mathsf{X}  = - \dfrac{\lambda}{2 \pi^2}\mathsf{S} + \frac{1}{3} \mathbf{1} +O(\lambda)^{-1/2}\,.
\end{align}
	One then finds
\begin{equation}
	\mathsf{D} = \frac{1}{\mathbf{1} - \mathsf{X}} = \left( \dfrac{\lambda}{2 \pi^2}\mathsf{S} + \frac{2}{3}\mathbf{1} \right)^{-1} + O(\lambda)^{-2}\,.
\end{equation}

We define new matrices $L$ and $\mathbb{D}$, different from the ones in \eqref{eq:LDdef}, 
\begin{equation}
	L_{k l} =  
	\begin{cases}
		1 & \text{if} ~k =  l~\\[4mm]
		-\dfrac{ E_k}{E_l}& \text{if} ~k= l+1\\[4mm] 
		0& \text{else}
	\end{cases},\qquad 	\mathbb{D}_{k l} =  
\begin{cases}
\dfrac{E_{k-1}}{E_k}  \dfrac{\lambda}{2 \pi^2}\dfrac{1}{4 k \sqrt{4 k^2-1} }& \text{if} ~k = l~,\\[4mm] 
0& \text{else}
\end{cases}
\end{equation}
where $E_k$ is expressed in terms of the modified Bessel function as
\begin{equation}
	E_k = \sqrt{2 k+1} \frac{ I_{2 k+1} \left( \sqrt{\frac{3\lambda}{2 \pi^2}} \right)}{  I_{1} \left( \sqrt{\frac{3\lambda}{2 \pi^2}} \right)}\,.
\end{equation} 

To proceed further we recall some properties of the modified Bessel function. A somewhat nonstandard recursion relation satisfied by the modified Bessel functions reads:
\begin{align}
	\label{eq:recursionBessel}
	\nu I_{2 \nu +3}(z) - \left( 1 + \nu (\nu +1) \dfrac{8}{z^2} \right) (2 \nu +1) I_{2 \nu+1}(z) 
	+ (\nu +1) I_{2 \nu -1}(z)=0
\end{align}
This can be proven by using the standard recursion relation 
\begin{equation}\label{eq:standardrecursion}
	\dfrac{2 \nu}{z} I_{\nu} (z) = I_{\nu-1}(z) -  I_{\nu+1}(z)\,.
\end{equation}
Using \eqref{eq:standardrecursion} on both terms in the right hand side of \eqref{eq:standardrecursion} gives
\begin{equation}
	\dfrac{2 \nu}{z^2 }I_{\nu} (z) = \dfrac{1}{2 (\nu-1)}I_{\nu-2}(z) -\dfrac{\nu}{\nu^2-1}I_{\nu}(z) +  \dfrac{1}{2 (\nu+1)}I_{\nu+2}(z) \,.
\end{equation}
This relation is equivalent to \eqref{eq:recursionBessel}, after renaming the order of $I_{\nu}(z)$.

Using the recursion relation of modified Bessel functions \eqref{eq:recursionBessel}, one can verify directly that
\begin{equation}
L^T \mathbb{D} L =   \dfrac{\lambda}{2 \pi^2}\mathsf{S} + \frac{2}{3}\mathbf{1}\,,
\end{equation}
and hence
\begin{align}
	L^{-1} \mathbb{D}^{-1} L^{-T} =  \left( \dfrac{\lambda}{2 \pi^2}\mathsf{S} + \frac{2}{3}\mathbf{1} \right)^{-1} = \mathsf{D} + O(\lambda)^{-2}\,.
\end{align}
Applying once again formula \eqref{ratiodet} we find:
\begin{align}\label{eq:0108b}
		1 + \Delta_k(\lambda) + O(\lambda)^{-2} =  \frac{\det \mathsf{D}_{(k)}}{\det \mathsf{D}_{(k-1)}} + O(\lambda)^{-2} = \mathbb{D}^{-1}_{kk}  + O(\lambda)^{-2}\,.
\end{align}

Plugging in the value of $\mathbb{D}_{kk}$ and expanding the modified Bessel functions for large $\lambda$, we obtain
\begin{equation}\label{eq:approx2}
	1 + \Delta_k(\lambda) = \dfrac{8\pi^2 k (2 k+1) }{\lambda} 
	- \frac{32 \sqrt{\frac{2}{3}} \pi ^3 k^2 (2 k+1)}{\lambda^{3/2}} +O(\lambda)^{-2}\,.
\end{equation}
Although the calculations above appear to be quite straightforward, it turns out that the coefficient $C_{3/2}$ of the $\lambda^{-3/2}$ term in the large $\lambda$ expansion of $\Delta_k(\lambda)$  calculated analytically here is different from the one obtained numerically, see \eqref{NikolayConjecture}:
\begin{equation}
	\begin{aligned}
	C_{3/2}^{\rm ana} &= - 32 \sqrt{\frac{2}{3}} \pi ^3 k^2 (2 k+1)\approx 810.1 k^2(2k+1),\\
	C_{3/2}^{\rm num} &= - 128\pi^2\log 2 k^2(2k+1) \approx 875.7k^2(2k+1)\,.\\
	\end{aligned}
\end{equation}
We believe that the reason for this discrepancy lies in the fact that the analytic calculation above is subtle and leads to a wrong result due to the fact that we have treated the infinite dimensional matrix $\mathsf{X} $ as a finite dimensional one. For example, first taking a large $\lambda$ expansion and then inverting an infinite matrix could potentially yield a different result than first inverting the matrix and then expanding at large $\lambda$. Also, we are calculating the large $\lambda$ expansion of the matrix $\mathsf{X}$ for each component separately. In doing this, perhaps one needs to be more careful to ensure that this expansion is uniform over all components of the matrix. While we have no rigorous proof that the analytic calculation above is wrong, due to the subtleties discussed here and the fact that the numerical results are very accurate and well-behaved we have chosen to trust the result of our numerical analysis and in particular the large $\lambda$ expansion in \eqref{NikolayConjecture}.

\section{Quadrature rules}\label{app:discretisation}

A quadrature rule for approximation to an integral is a set of points $x_i$ and weights $w_i$ such that
\begin{equation}
I \equiv \int_a^b dx f(x) \approx \sum_{i=1}^n w_i f(x_i) \equiv I_n\,.
\end{equation}
The quadrature rules we describe in this section converge exponentially fast. Namely, if $f(x)$ can be extended to an analytical function $f(z)$ in a region around the interval $\left[a,b\right]$ then 
\begin{equation}
 | I - I_n| \le e^{- c n}\,,
\end{equation}
with $c$ a positive constant. Typically, the larger the region of analyticity, the larger $c$. The reason we want fast quadrature rules is that $n$ should not be too large because we solve linear algebra problems
with matrices of size $n \times n$, see equation \eqref{eq:eq12}. The memory requirement will then be of order $O(n)^2$, and the timing of order $O(n)^3$, so this scales quite badly with $n$.

\subsection*{Gauss-Laguerre}
This is a quadrature scheme for the interval $[0, +\infty[$. It is the Gaussian quadrature scheme based on Laguerre polynomials. The weights $w_i$ and points $x_i$ with $i = 1 , \ldots, n$ are such that all polynomials of degree not greater than $ 2 n -1$ are integrated exactly. More specifically, if
\begin{equation}
I \equiv \int_0^{+\infty}\! dx\ e^{-x} f(x),\qquad I_n \equiv \sum_{i=1}^n w_i f(x_i)\,,
\end{equation}
then $I = I_n$  for all polynomials $f(x)$ of degree not greater than $ 2 n -1$. Initially, this seemed to be a natural quadrature scheme to use in our setup because the function $W(t)$ in \eqref{defW} decays exponentially fast for $t \to +\infty$. Also, although the weights $w_i$ and points $t_i$  are not directly available in {\tt Mathematica}, they can be calculated efficiently with the Golub-Welsch algorithm. This quadrature scheme works quite well for $\lambda \lesssim 100$, however for larger values of $\lambda$ we need to use larger values of $n$ ($n \gtrsim 100$). For large values of $n$, it turns out that many weights are very small ($\sim 10^{-100}$), so they are in a sense wasted. This is also documented in the literature, see Sections 5 and 6 in \cite{Tref}. Therefore, it is better to use a different quadrature scheme.

\subsection*{Quadrature rules on the interval $[-1 ,1]$}

Some well-known quadrature rules are: 
\begin{description}
	\item[Gauss-Legendre] Here $x_i$ are given by the zeros of Legendre polynomials. These are actually points in the interval $[0,1]$, but a linear change of variable converts these to the interval  $[-1 ,1]$.
	\item[Clenshaw-Curtis] Here $x_k = \cos\left(\frac{k \pi}{n}\right)$ with $k=0,1, \ldots, n$.
	\item[Fej\'er type 1] Here $x_k = \cos\left(\frac{(k - \frac{1}{2}) \pi}{n}\right)$ with $k=1,2, \ldots, n$. 
\end{description}

Gauss-Legendre is perhaps theoretically nicer than 	Clenshaw-Curtis, but the advantage of Clenshaw-Curtis is that its weights and points are much easier to compute. Also, {\tt Mathematica} has a function to compute the weights and points of Clenshaw-Curtis. However, the end points $x_0=1$ and $x_n = -1$ are part of the quadrature points. In the integral equation we want to solve, we prefer not to use the left end point. Therefore, we settled on Fej\'er type 1.
\subsection*{Fej\'er type 1}
The quadrature points are:
\begin{equation}\label{eq:Fejertheta}
x_k = \cos \theta_k,\quad\text{with}\quad 	\theta_k = (2 k -1) \frac{\pi}{2 n}, \quad k = 1 , 2, \ldots, n\,,
\end{equation}
and the weights $w_k$ can for example be found in \cite{Waldvogel}\footnote{If the reader wants to check formula \eqref{eq:Fejerweights}, we found Fej\'er's own calculation very readable \cite{Fejer}.}
\begin{equation}\label{eq:Fejerweights}
	w_k = \frac{2}{n} \left[ 1 - 2 \sum_{r =1}^{ \frac{n-1}{2} } \frac{\cos(2 r \theta_k)}{4 r^2 -1} \right]
\end{equation}

 The points $x_k$ and weights $w_k$ are not directly available in {\tt Mathematica}. However, there is code available at Wolfram Function Repository  which can be used for their calculation \cite{WolframResource}.\footnote{The function is called \texttt{FejerQuadratureWeights} and is based on the Fast Fourier Transform. We checked for many cases that \texttt{FejerQuadratureWeights} indeed produces $x_i$ and $w_i$ that are numerically the same as the ones given by \eqref{eq:Fejertheta} and \eqref{eq:Fejerweights}.}
\subsection*{Error analysis}
To integrate numerically over the interval $[0,+\infty[$ we truncate the interval to  $[0,L]$ and then use  Fej\'er type 1 on the truncated interval.
This amounts to the approximation
\begin{equation}
\int_0^{+\infty}\!\!\! dt\ \ f(t) \approx \int_0^L \!\!\! dt\ \ f(t) \approx \sum_{i=1}^n w_i f(x_i)\,.
\end{equation}
This procedure leads to two sources of error: firstly, there is the truncation error which goes to zero if $L \to +\infty$; secondly, there is the discretization error which goes to zero if $n \to +\infty$.
One possibility is to keep $L$ and $n$ independent, and test for accuracy by increasing both $L$ and $n$ separately. Another possibility is to relate $L$ and $n$ in such a way that the truncation error is roughly equal to the discretization error. We observed that the scaling $L \sim n^{2/3}$ works well in practice. 

It is important to note that if $n \to \infty$ then $L \to \infty$, so the truncation error goes to zero. Also if $n \to \infty$, the discretization size $h = \frac{L}{n} \sim n^{-1/3} \to 0$, so the discretization error will go to zero as well.

\section{Degenerate kernels}
\label{sec:degkernels}

In this section we collect some formulae for integral operators with degenerate kernel. These formulae are well-known but included here for convenience.
The integral operator
\begin{equation}\label{eq:deg1}
	f(x) \mapsto f(x) + \int_{\alpha}^{\beta}\!\!\!\! dy\ K(x,y) f(y) \,,
\end{equation}
has degenerate kernel (also known as kernel of finite rank or separable kernel) if $K(x,y)$ can be expressed as the finite sum\footnote{We do not use the summation convention for repeated indices in this section.}
\begin{equation}\label{eq:deg2}
	K(x,y) = \sum_{i,j=1}^n a_i(x)\ C_{ij}\ b_j(y)\,.
\end{equation}
Integral equations with degenerate kernel can be solved in closed form as follows. Suppose the integral equation is
\begin{equation}\label{eq:deg3}
	f(x) + \int_{\alpha}^{\beta}\!\!\!\! dy\ K(x,y) f(y) = g(x) \,,
\end{equation}
with $K(x,y)$ as in \eqref{eq:deg2}. Define $f_i =  \int_{\alpha}^{\beta}\!\! dy\ b_i(y) f(y)$, then equation \eqref{eq:deg3} is
\begin{equation}\label{eq:deg4}
	f(x) + \sum_{i,j=1}^n a_i(x) \ C_{ij} f_j = g(x) \,.
\end{equation}
Multiplying with $b_k(x)$ and integrating over $x$ gives
\begin{equation}\label{eq:deg5}
	f_k + \sum_{i,j=1}^n A_{ki} C_{ij} f_j =g_k \,,
\end{equation}
where
\begin{equation}\label{eq:deg6}
	A_{ki} = \int_{\alpha}^{\beta}\!\!\!\! dx\ b_k(x) a_i(x) \,,
\end{equation}
and $g_k =  \int_{\alpha}^{\beta}\!\! dx\ b_k(x) g(x)$. 
The system of $n$ linear equations \eqref{eq:deg5} can be solved exactly for $f_k$. Inserting this solution in \eqref{eq:deg4} gives\footnote{$M_{ij}^{-1}$ is the $ij$ component of the matrix $M^{-1}$.}
\begin{equation}\label{eq:deg7}
	f(x) = g(x) - \sum_{i,j,l=1}^n a_i(x) C_{ij} (1 + AC)_{jl}^{-1} g_l \,.
\end{equation}
The solution of \eqref{eq:deg3} is thus
\begin{equation}\label{eq:deg8}
	f(x) = g(x) -  \int_{\alpha}^{\beta}\!\!\!\! dy\ L(x,y) g(y) \,,
\end{equation}
with 
\begin{equation}\label{eq:deg9}
	L(x,y) = \sum_{i,j=1}^n a_i(x) [ C ( 1 + AC)^{-1}]_{ij} b_j(y) \,.
\end{equation}
In the literature $L(x,y)$ is called the resolvent.

The Fredholm determinant of an integral operator with degenerate kernel can also be calculated analytically. Here is a derivation of the formula.
We discretize the integral with discretisation points $x_{\mu}$ and weights $w_{\mu}$ with $\mu, \nu = 1,2, \ldots, m$. Then the Fredholm determinant is equal to
\begin{equation}
\det(1 + K) = \lim_{m \to \infty} \det_{\mu,\nu=1}^m ( \delta_{ \mu \nu} + K(x_{\mu} , x_{\nu}) w_{\nu})\,.
\end{equation}
Write
\begin{equation}
K(x_{\mu} , x_{\nu}) w_{\nu} = \sum_{i,j=1}^n a_i(x_{\mu})\ C_{ij}\ b_j( x_{\nu}) w_{\nu} = ( M C N )_{\mu\nu}\,,
\end{equation}
with $M_{\mu i} = a_i(x_{\mu})$ and $N_{j \nu} =  b_j( x_{\nu}) w_{\nu}$. Then
\begin{equation}
\det(\mathbf{1}_{m \times m } + MCN) = \det(\mathbf{1}_{n \times n } + CNM)\,.
\end{equation}
Since
\begin{equation}
( CNM)_{ij} = \sum_{k=1}^n \sum_{\nu=1}^m C_{ik} b_k(x_{\nu}) w_{\nu} a_j(x_{\nu})\,,
\end{equation}
we find that in the limit $m \to \infty$
\begin{equation}
( CNM)_{ij} = \sum_{k=1}^n C_{ik} A_{kj}\,,
\end{equation}
with $A_{kj}$ defined in \eqref{eq:deg6}.
Alltogether, one has
\begin{equation}\label{eq:deg10}
	\det(\mathbf{1} + K) = \det(\mathbf{1}_{n \times n } + CA) \,.
\end{equation}
This is a closed form formula for the Fredholm determinant because the determinant on the right hand side is of a finite $n \times n$ matrix.

\section{Numerical data}\label{app:tables}

Here we provide two tables with some of our numerical data for $\Delta_k(\lambda)$ for $k=1,2,3$ as well as for the free energy $\mathcal{F}$.
\begin{table}[H]
$$
\begin{array}{|c||c|c|c|}
	\hline
	\lambda  & 1+\Delta_1(\lambda) &1+\Delta_2(\lambda) &1+\Delta_3(\lambda) \\
	\hline 
	\hline
	e^3 & 0.967087942591236   (00\! \pm \! 11 ) & 0.998182 &
	0.999886471522795   (00\! \pm \! 11 ) \\
	e^4 & 0.84413354083007   (60\! \pm \! 19 ) & 0.976408560611106 
	(00\! \pm \! 22 ) & 0.996135433920974   (00\! \pm \! 11 ) \\
	e^5 & 0.591355036026568   (60\! \pm \! 22 ) & 0.867655885531129 
	(00\! \pm \! 11 ) & 0.95770388339493   (20\! \pm \! 22 ) \\
	e^6 & 0.3263339352130   (10\! \pm \! 12 ) & 0.62836903552457  
	(64\! \pm \! 11 ) & 0.80851244411776   (50\! \pm \! 30 ) \\
	e^7 & 0.1524442716515   (11\! \pm \! 10 ) & 0.36148255438002  
	(6\! \pm \! 5 ) & 0.547853721918213   (5\! \pm \! 8 ) \\
	e^8 & 0.064525566747991   (9\! \pm \! 9 ) & 0.1749433084896  
	(77\! \pm \! 15 ) & 0.2997641417256   (9\! \pm \! 5 ) \\
	e^9 & 0.02579506509172   (98\! \pm \! 16 ) & 0.07589996453754  
	(0\! \pm \! 7 ) & 0.1407940457851   (2\! \pm \! 4 ) \\
	e^{10} & 0.00997266161958   (43\! \pm \! 27 ) & 0.0308295464823 
	(8\! \pm \! 6 ) & 0.0600521108447   (98\! \pm \! 30 ) \\
	e^{11} & 0.003779867460756   (4\! \pm \! 4 ) & 0.0120383586528 
	(52\! \pm \! 14 ) & 0.0241553127444   (05\! \pm \! 15 ) \\
	e^{12} & 0.001415785217   (9\! \pm \! 4 ) & 0.00459090014  
	(08\! \pm \! 27 ) & 0.00937868572   (6\! \pm \! 8 ) \\
	e^{13} & 0.0005265339   (42\! \pm \! 19 ) & 0.001726033   (07\!
	\pm \! 12 ) & 0.003564618   (8\! \pm \! 4 ) \\
	e^{14} & 0.000194980   (46\! \pm \! 21 ) & 0.00064338   (77\!
	\pm \! 14 ) & 0.00133750   (4\! \pm \! 4 ) \\
	\hline
\end{array}
$$
\caption{$\Delta_k(\lambda)$ with $k=1,2,3$ for some values of $\lambda$. These values are calculated with the Nystr\"om method explained in Section \ref{sec:Deltanum}. We have used different values of $L$ and $m$ to estimate the accuracy of the numerical values.}
\end{table}
\begin{table}[H]
$$
\begin{array}{|l|l||l|l|}
	\hline
	\lambda  & {\cal F} &\lambda  & {\cal F}\\
	\hline 
	\hline
1 & 9.3017 \times 10^{-6} & 100 & 0.227879 \\
2 & 0.0000663066 & 200 & 0.50639 \\
3 & 0.000200935 & 300 & 0.759961 \\
4 & 0.000430494 & 400 & 0.991783 \\
5 & 0.000764363 & 500 & 1.20637 \\
6 & 0.00120686 & 600 & 1.40708 \\
7 & 0.00175905 & 700 & 1.59638 \\
8 & 0.00241988 & 800 & 1.77606 \\
9 & 0.00318697 & 900 & 1.94753 \\
\hline
10 & 0.00405711 & 1000 & 2.11184 \\
20 & 0.0175565 & 2000 & 3.49445 \\
30 & 0.0373082 & 3000 & 4.60102 \\
40 & 0.060823 & 4000 & 5.55364 \\
50 & 0.0866247 & 5000 & 6.40401 \\
60 & 0.11382 & 6000 & 7.17993 \\
70 & 0.141848 & 7000 & 7.89843 \\
80 & 0.170347 & 8000 & 8.57086 \\
90 & 0.199078 & 9000 & 9.20523 \\
	\hline
\end{array}
$$
\caption{The free energy ${\cal F}$ for some values of $\lambda$. These values are calculated with the Bornemann method explained in Section \ref{sec:Fnum}. We have used appropriate settings of $L$ and $m$ to ensure that all printed digits are correct.}
\end{table}

%



\end{document}